\documentclass[comsoc,onecolumn,draftclsnofoot,12pt]{IEEEtran}
\usepackage[T1]{fontenc}
\usepackage[cmintegrals]{newtxmath}

\usepackage{cite}

\ifCLASSINFOpdf
\else
\fi
\usepackage{changebar}

\usepackage{tikz}
\usetikzlibrary{math}
\usetikzlibrary{shapes.geometric,calc}
\usetikzlibrary{arrows.meta, quotes, angles}
\usetikzlibrary{decorations}
\usetikzlibrary{chains,shapes.misc,positioning,scopes}
\usepackage{circuitikz}
\usetikzlibrary{circuits.ee.IEC}
\usetikzlibrary{decorations}
\usetikzlibrary{dsp}
\usetikzlibrary{shadows}
\tikzset{
diagonal fill/.style 2 args={fill=#2, path picture={
\fill[#1, sharp corners] (path picture bounding box.south west) -|
                         (path picture bounding box.north east) -- cycle;}},
reversed diagonal fill/.style 2 args={fill=#2, path picture={
\fill[#1, sharp corners] (path picture bounding box.north west) |- 
                         (path picture bounding box.south east) -- cycle;}}
}

\usepackage{amsmath}
\usepackage{trsym}

\usepackage{nicefrac}
\usepackage{siunitx}

\usepackage{cite}

\usepackage[cmintegrals]{newtxmath}
\usepackage{bm}

\usepackage{caption}
\usepackage{subcaption} 

\usepackage[linesnumbered,ruled]{algorithm2e}
\usepackage{colortbl}

\newcommand{\Kprim}{\bm{K}}
\newcommand{\Kadj}{\tilde{\bm{K}}}

\newcommand{\her}{^{\scriptscriptstyle\mathrm{H}}}
\newcommand{\tran}{^{\scriptscriptstyle\mathrm{T}}}
\newcommand{\dint}[1]{\,\mathrm{d}#1}

\newcommand{\expE}[1]{\mathrm{e}^{#1}}
\newcommand{\jcomp}{\mathrm{j}}

\newcommand{\x}{\bm{x}}

\newcommand{\linpr}{\left\langle}
\newcommand{\rinpr}{\right\rangle}

\newcommand{\Lp}{\bm{\mathrm{L}}}

\newcommand{\As}{\bm{\mathcal{A}}}

\newcommand{\Cs}{\bm{\mathcal{C}}}

\newcommand{\Ks}{\bm{\mathcal{K}}}

\begin{document}

%



\title{Transfer Function Models for Cylindrical MC Channels with Diffusion and Laminar Flow}

%
%
%

\author{Maximilian Sch\"afer,
        Wayan~Wicke,
        Lukas~Brand,        
        Rudolf Rabenstein,
        and~Robert~Schober
\vspace*{-1.5cm}
}

\maketitle

\begin{abstract}
\vspace*{-2ex}
The analysis and design of advection-diffusion based molecular communication (MC) systems in cylindrical environments is of particular interest for applications such as micro-fluidics and targeted drug delivery in blood vessels. Therefore, the accurate modeling of the corresponding MC channel is of high importance. 
The propagation of particles in these systems is caused by a combination of diffusion and flow with a parabolic velocity profile, i.e., laminar flow. 
The propagation characteristics of the particles can be categorized into three different regimes: 
The flow dominant regime where the influence of diffusion on the particle transport is negligible, the dispersive regime where diffusion has a much stronger impact than flow, and the mixed regime where both effects are important. 
For the limiting regimes, i.e., the flow dominant and dispersive regimes, there are well-known solutions and approximations for particle transport. 
In contrast, there is no general analytical solution for the mixed regime, and instead, approximations, numerical techniques, and particle based simulations have been employed. 
In this paper, we develop a general model for the advection-diffusion problem in cylindrical environments which provides an analytical solution applicable in all regimes. The modeling procedure is based on a transfer function approach and the main focus lies on the incorporation of laminar flow into the analytical model. 
The properties of the proposed model are analyzed by numerical evaluation for different scenarios including the uniform and point release of particles. We provide a comparison with particle based simulations and the well-known solutions for the limiting regimes to demonstrate the validity of the proposed analytical model.
\vspace*{-1ex}
\end{abstract}


%
\IEEEpeerreviewmaketitle

\section{Introduction}
\label{sec:intro}

\IEEEPARstart{R}{ecently}, the application of communication engineering principles to biomedical problems has spawned the emerging interdisciplinary research field of molecular communication (MC). 
Comprehensive descriptions of MC can be found in \cite{Nakano_et_al:Book:2013,Farsad_et_al:IEEECommSurvTutorials:2016}, while a tutorial review of theoretical concepts and modeling techniques is provided in \cite{jamali:ieee:2019}.
%
MC is ubiquitous in natural biological systems and has a high potential for bio-medical applications such as targeted drug delivery, health monitoring \cite{Farsad_et_al:IEEECommSurvTutorials:2016,Felicetti_Modeling_2014,Farokhzad:acs:2009}, and micro-fluidic channel design \cite{Bicen_etal:IEEETrSigProc:2013}. 
Besides medical applications, MC may be applied in industrial settings, e.g., for monitoring of chemical reactors and pipelines \cite{Lluis:cn:2009}.
The main difference between MC and classical communications is the means of transport of information from the transmitter (TX) to the receiver (RX). 
While classical communication systems rely on transport by electro-magnetic or acoustic wave propagation, mostly in free space, motivated by biological systems several different transport mechanisms have been considered for MC.
These mechanisms include diffusion, gap-junction, and molecular motor based transport \cite{jamali:ieee:2019}. 
	%
In fluid environments, diffusion often occurs together with advection, which is prevalent, e.g., in blood vessels or pipelines. 
In this case, the particles are diffusing randomly and are additionally affected by a background flow. The flow in blood vessels and pipelines is characterized as Poiseuille flow, which exhibits a specific laminar flow profile with a radial dependence of the flow velocity \cite{jamali:ieee:2019}.
	
The accurate modeling of MC channels is crucial for the analysis of naturally occurring MC systems and the design of artificial MC systems.  
As the analysis and design of advection-diffusion based MC systems in cylindrical environments is of particular interest, e.g., for micro-fluidic applications and targeted drug delivery systems, corresponding models have been extensively studied \cite{Zoofaghari:ieee:2019,schaefer:icc:2019, unterweger:2018, Dinc:ieee:2019,Lo:ieee:2019, kuscu_modeling_2018,He_Channel_2016,wicke:globecom:2018 , Aris:1956, Probstein:Book:2005, Chahibi:ieee:2015, Bicen_etal:IEEETrSigProc:2013}.
Hereby, the most challenging aspect of the modeling is the correct incorporation of the parabolic flow profile which introduces a coupling between the axial and cross-sectional particle distributions.
Therefore, many existing models resort to the common plug-flow simplification, which assumes a uniform axial flow in the cylinder \cite{Levenspiel_Chemical_1999}. 
Based on Green's functions, the authors of \cite{Zoofaghari:ieee:2019} present an MC channel model for a cylindrical environment with plug flow, a first-order degradation reaction, and partially absorbing boundaries. 
Advection and diffusion of magnetic nano-particles in a cylinder is considered in \cite{schaefer:icc:2019, unterweger:2018}, where the particles are also affected by an external magnetic force.  

There are only a few analytical models in the MC literature which consider Poiseuille flow. 
In \cite{Dinc:ieee:2019}, the impulse response of a three-dimensional (3D) advection-diffusion channel is derived by approximating the laminar flow profile by a piecewise function for the axial distribution of particles. 
A Markovian-based channel model is presented in \cite{Lo:ieee:2019}, where the cross section of the cylinder is divided into rings and the laminar flow profile is approximated by the mean of the flow velocity in each ring.
A heuristic parametric model is proposed in \cite{kuscu_modeling_2018} for micro-fluidic MC channels with surface-based receivers.
For modeling the influence of Poiseuille flow on the propagation of particles, it is convenient to categorize the transport process into three different regimes, namely the flow dominant, dispersive, and mixed regimes \cite[Sec.~II-B]{jamali:ieee:2019}, \cite{Probstein:Book:2005}.  
In \cite{wicke:globecom:2018}, an analytical model for the flow dominant regime is presented for both uniform and point release of particles. 
The effect of diffusion is neglected in this regime. 
Dispersion is used in \cite{Aris:1956} to model the interplay of diffusion and laminar flow, which is also known as Taylor dispersion where an effective diffusion coefficient is utilized together with a plug flow approximation \cite{Probstein:Book:2005}. 
The resulting model is applicable in the dispersive regime, where the interaction of diffusion and laminar flow yields a uniform distribution of particles in the cross section \cite{wicke:globecom:2018}. 
This approximation for particle propagation	is applied for the modeling of MC channels in, e.g., \cite{wicke:globecom:2018, He_Channel_2016, Chahibi:ieee:2015,Bicen_etal:IEEETrSigProc:2013}. 
In the mixed regime, the particles are affected by both diffusion and laminar flow and neither is negligible. 
Therefore, the solutions for the flow dominant and dispersive regimes are not applicable and either numerical techniques or the simplified models in \cite{Dinc:ieee:2019,Lo:ieee:2019,kuscu_modeling_2018} have been employed.
To the best of the authors' knowledge, a general analytical model for cylindrical MC channels with diffusion and laminar flow, which is applicable in all three regimes, has not been reported, yet. 

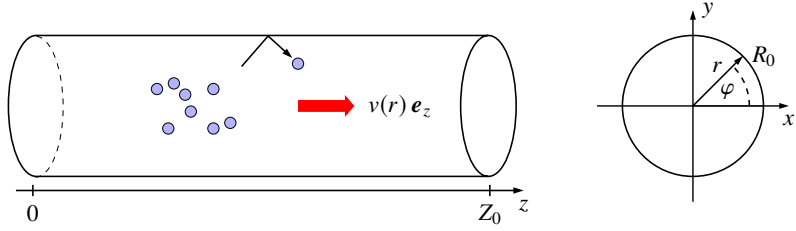
\begin{figure*}[t]

\centering
\scalebox{0.75}{
\begin{tikzpicture}
	
	\node[cylinder,draw=black,thick,aspect=3.5,minimum height=9cm,minimum width=2.5cm,shape border rotate=0] (A) {};
	
	\draw[dashed]
	let \p1 = ($ (A.after bottom) - (A.before bottom) $),
	\n1 = {0.5*veclen(\x1,\y1)-\pgflinewidth},
	\p2 = ($ (A.bottom) - (A.after bottom)!.5!(A.before bottom) $),
	\n2 = {veclen(\x2,\y2)-\pgflinewidth}
	in
	([xshift=-\pgflinewidth] A.before bottom) arc [thick, start angle=270, delta angle=180,
	x radius=\n2, y radius=\n1];
	
	\node[circle,inner sep=0pt, minimum size=0.2cm, draw,fill=blue!30](part) at (-1,0.2){};
	\node[circle,inner sep=0pt, minimum size=0.2cm, draw,fill=blue!30](part) at (-1.2,0.4){};
	\node[circle,inner sep=0pt, minimum size=0.2cm, draw,fill=blue!30](part) at (-1.5,0.3){};
	\node[circle,inner sep=0pt, minimum size=0.2cm, draw,fill=blue!30](part) at (-0.5,-0.4){};
	\node[circle,inner sep=0pt, minimum size=0.2cm, draw,fill=blue!30](part) at (-0.2,-0.3){};
	\node[circle,inner sep=0pt, minimum size=0.2cm, draw,fill=blue!30](part) at (-0.9,-0.1){};
	\node[circle,inner sep=0pt, minimum size=0.2cm, draw,fill=blue!30](part) at (-1.3,-0.4){};
	\node[circle,inner sep=0pt, minimum size=0.2cm, draw,fill=blue!30](part) at (-0.5,0.3){};
	
	\draw[line width=6pt,-{Triangle[width=1.8*6pt,length=0.8*6pt]}, draw=red](1,0) -- (2, 0)node[right]{$v(r)\,\bm{e}_z$};
	\draw[dspconn,thick] (-4,-1.5) to (5,-1.5) node[below]{$z$};
	\draw[dspline,thick] (4.4,-1.4) to (4.4,-1.6) node[below]{$Z_0$};
	\draw[dspline,thick] (-3.7,-1.4) to (-3.7,-1.6) node[below]{$0$};
	
	
	\draw[dspconn, thick] (0,0.7) -- (0.47, 1.24) --
	(0.9, 0.85);
	\node[circle,inner sep=0pt, minimum size=0.2cm, draw,fill=blue!30](part) at (1,0.75){};

	\draw[thick] (8,0) circle (1.25cm);
	\draw[dspconn] (8,-1.7) -- (8, 1.7)node[right]{$y$};
	\draw[dspconn] (6.3,0) -- (9.7, 0)node[below]{$x$};
	
	\draw[dspconn] (8,0) -- node[midway, above]{$r$}(8.8839,0.8839) node[right]{$R_0$};
	
	\draw
    (9.25,0) coordinate (a) 
    -- (8,0) coordinate (b) 
    -- (8.8839,0.8839) coordinate (c)
 pic["$\varphi$"{shift={(-.5cm, -.2cm)}}, draw=black, dashed, thick, angle eccentricity=1.2, angle radius=1cm]
    {angle=a--b--c};
	
	\end{tikzpicture}
	}
	\vspace{-2ex}
    \caption{\small Diffusing particles (blue circles) in a cylinder of volume $V$ subject to horizontal laminar flow $v(r)$. The cylinder has radius $R_0$, length $Z_0$, and the boundaries are fully reflective.}
    \label{fig:1}
    \vspace*{-3ex}
\end{figure*}

In this paper, we establish a general analytical model for the transport of particles by diffusion and laminar flow in cylindrical MC channels, see Fig.~\ref{fig:1}.
The starting point for the modeling is the well-known advection-diffusion equation, a partial differential equation (PDE). 
Subsequently, a transfer function model (TFM) is established.
The TFM approach
is based on the modal expansion of a PDE into a set of eigenfunctions and eigenvalues, and provides a representation of the problem in a spatio-temporal transform domain \cite{churchill:1972,Curtain-Zwart:infdimsysth:1995}. 
Finally, the solution of the PDE is represented as the output of a state-space description (SSD) and in terms of a concentration Green's function (CGF) \cite{rabenstein:ijc:2017}. 
The TFM approach has been applied for the modeling of cylindrical and spherical MC systems \cite{schaefer:icc:2019, schaefer:icc:2020}, where it has been used to realize complex boundary conditions. 
However, laminar flow was not considered in \cite{schaefer:icc:2019, schaefer:icc:2020}.   
Therefore, in this paper, the TFM approach is extended to incorporate the influence of laminar flow.  
%
%
%
%
%
To this end, the PDE is first reduced to a simple diffusion equation which is solved in terms of an SSD of an open loop system. 
Then, the influence of laminar flow is incorporated via a feedback system that is attached to the open loop SSD to form a closed loop SSD. 
The design of feedback systems is well known in control theory, see e.g., \cite{Deutscher:book:2012}. 
Here, this approach is adopted to incorporate the influence of laminar flow. 
%
%
The main contributions of this paper can be summarized as follows:
\begin{itemize}
	\item We derive a general analytical model for the transport of particles by diffusion and laminar flow in cylindrical MC channels, which is applicable in all three particle propagation regimes.  
	\item The proposed model can be formulated either in terms of a CGF for analytical analysis or an SSD for efficient numerical evaluation. 
	\item Uniform and point release of the particles are considered for analysis. Comparisons with results from particle based simulations (PBS) and known solutions for the flow dominant \cite{wicke:globecom:2018} and dispersive regimes \cite{Aris:1956} verify the validity of the proposed model.
\end{itemize}
%
The remainder of this paper is structured as follows: Section~\ref{sec:math} presents the considered advection-diffusion problem and introduces its mathematical description. 
Section~\ref{sec:tfm1} establishes a TFM of the 3D diffusion process, i.e., the open loop SSD. 
In Section~\ref{sec:tfm2}, the influence of laminar flow is incorporated via a feedback system that is attached to the open loop SSD to form a closed loop SSD. 
The validity of the derived model is verified in Section~\ref{sec:simul} via numerical evaluation. 
Section~\ref{sec:analysis} further analyses the proposed model and discusses its practical implementation. Finally, Section~\ref{sec:conc} concludes the paper and presents several topics for further research.

\section{System Model and Mathematical Preliminaries}
\label{sec:math}

\subsection{System Model}
\vspace*{-1ex}
The cylindrical volume $V$ shown in Fig.~\ref{fig:1} can be characterized by vector $\bm{x} = \left[r, \varphi, z\right]$ in cylindrical coordinates and its radial boundary $\partial V_\mathrm{r}$ and axial boundary $\partial V_\mathrm{z}$ as follows
\begin{align}
V &= \{\bm{x} = \left[r, \varphi, z\right]\tran \big\vert\, 0 \le r \le R_0,\, -\pi \le \varphi \le \pi,\, 0 \le z \le Z_0\}, \label{eq:1}\\
\partial V_\mathrm{r} &= \{\bm{x} = \left[r, \varphi, z\right]\tran \big\vert\, r = R_0, \, -\pi \le \varphi \le \pi,\, 0 \le z \le Z_0\}, \label{eq:2}\\
\partial V_\mathrm{z} &= \{\bm{x} = \left[r, \varphi, z\right]\tran \big\vert\, 0 \le r \le R_0, \, -\pi \le \varphi \le \pi,\, z = 0,\,Z_0\}.\label{eq:3}
\end{align}
The diffusion and flow of particles in $V$ are described by an initial-boundary value problem (IBVP) in terms of the particle concentration $p(\bm{x},t)$ in $\si{\mole\per\meter^3}$ and the vector of particle flux $\bm{i}(\bm{x},t)$ in $\si{\mole\per\square\meter\per\second}$. 
The IBVP consists of a set of PDEs defined on \eqref{eq:1}, a set of boundary conditions (BCs) defined on \eqref{eq:2}, \eqref{eq:3}, and a set of initial conditions (ICs) defined on \eqref{eq:1} for $t = 0$. 
The PDE that describes the particle concentration $p(\bm{x},t)$ in volume $V$ under the influence of diffusion and flow is the advection-diffusion equation \cite[Eq.~(5.22)]{Bruus:2007}
\begin{align}
\frac{\partial}{\partial t}p(\bm{x},t) = D\,\mbox{div}\left(\mbox{grad}\left( p(\bm{x},t)\right)\right) - \mbox{div} \left(p(\bm{x},t) \bm{v}(\bm{x})\right),
\label{eq:4}
\end{align}
where $\nicefrac{\partial}{\partial t}$ denotes the partial derivative with respect to time and the operators $\mbox{div}\left(\cdot\right)$ and $\mbox{grad}\left(\cdot\right)$ denote the divergence and gradient operators in cylindrical coordinates, respectively. Constant $D$ is the diffusion coefficient in $\si{\meter\square\per\second}$ and $\bm{v}(\bm{x})$ is the velocity vector. 
For the case that the considered scenario in Fig.~\ref{fig:1} represents a straight channel with no-slip boundary conditions, the velocity profile is referred to as Poiseuille flow. Assuming that the channel in Fig.~\ref{fig:1} contains a Newtonian fluid with viscosity $\eta$, the velocity vector simplifies to a radius dependent laminar flow velocity  \cite[Ch.~3]{Bruus:2007}
\vspace*{-2ex}
\begin{align}
&\bm{v}(\bm{x}) = v(r)\,\bm{e}_z,
&v(r) = v_0\left(1 - \frac{r^2}{R_0^2}\right) \label{eq:5},
\end{align}
where $\bm{e}_z$ is the unit vector in $z$-direction and $v_0 = 2\,v_\mathrm{eff}$, while $v_\mathrm{eff}$ is the mean velocity in the channel. 
Decomposing the PDE in \eqref{eq:4} into a continuity equation and a concentration gradient equation and exploiting \eqref{eq:5} yields a set of two PDEs describing the dynamics of particle concentration $p(\bm{x},t)$ and flux $\bm{i}(\bm{x},t)$ in volume \eqref{eq:1} as follows
\begin{align}
\frac{\partial}{\partial t} p(\bm{x},t) + \mbox{div}\left( \bm{i}(\bm{x},t)\right) & = f_\mathrm{s}(\bm{x},t), &&\bm{x}\in V, 0 < t \le \infty, \label{eq:6}\\
\bm{i}(\bm{x},t) + D\,\mbox{grad}\left( p(\bm{x},t)\right) - v(r)\, p(\bm{x},t)\, \bm{e}_z &= 0, && \bm{x}\in V, 0 < t \le \infty, \label{eq:7}
\end{align}
where the vector of fluxes $\bm{i}\in\mathbb{R}^{3\times 1}$ contains the components of the three coordinate directions
\begin{align}
\bm{i}(\bm{x},t) = \begin{bmatrix}
i_r(\bm{x},t) & i_\varphi(\bm{x},t) & i_z(\bm{x},t)
\end{bmatrix}\tran \label{eq:8}
\end{align}
with $(\cdot)\tran$ denoting transposition.
Function $f_\mathrm{s}$ in \eqref{eq:6} denotes a space and time-dependent source function that can be used to model particle injection into the channel.
In addition to PDEs \eqref{eq:6} and \eqref{eq:7}, a set of boundary and initial conditions is defined
\begin{align}
p(\bm{x},t)\big\vert_{z = 0, Z_0} &= 0, && \bm{x}\in \partial V_\mathrm{z}, \, 0 < t \le \infty, \label{eq:9}\\
i_r(\bm{x},t)\big\vert_{r = R_0} &= 0, &&\bm{x}\in \partial V_\mathrm{r}, \, 0 < t \le \infty,\label{eq:10}\\
p(\bm{x},t)\big\vert_{t = 0} &= p_\mathrm{init}(\bm{x}), &&\bm{x}\in V,\, t = 0. \label{eq:11}
\end{align}
The boundary conditions of the cylinder in $z$-direction \eqref{eq:9} imply a cylinder with absorbing boundary, i.e., particles can leave the cylinder at $z = 0, Z_0$. 
We note that mostly channels of infinite length have been considered in the literature, e.g., \cite{schaefer:icc:2019,Zoofaghari:ieee:2019}. 
However, due to the applied modeling approach (see Section~\ref{sec:tfm1}) a bounded domain has to be chosen.  
%
%
The radial boundaries of the cylinder are fully reflective \eqref{eq:10}, and therefore the particle flux $i_r$ in radial direction is zero on $\partial V_\mathrm{r}$. The initial distribution of the particles $p_\mathrm{init}$ in $V$ is defined by the IC \eqref{eq:11}.

\subsection{Vector Formulation}
\vspace*{-1ex}
For the application of the proposed modeling approach in Section~\ref{sec:tfm1}, PDEs \eqref{eq:6}, \eqref{eq:7} and initial conditions \eqref{eq:11} are reformulated in a unified vector formulation as follows \cite{rabenstein:ijc:2017}
\begin{align}
\left[\frac{\partial}{\partial t}\bm{D} - \Lp \right]\bm{y}(\bm{x},t) &= \bm{f}_\mathrm{e}(\bm{x},t) + \bm{v}_\mathrm{flow}(\bm{x},t), 
&&\bm{x}\in V, 0 < t \le \infty, \label{eq:12}\\
\Lp &= \bm{A} + \nabla\bm{B}, \label{eq:13}\\
\bm{y}(\bm{x},t)\big\vert_{t = 0} &= \bm{y}_\mathrm{init}(\bm{x}), &&\bm{x}\in V,\, t = 0, \label{eq:14}
\end{align}
where the vector of variables $\bm{y} \in \mathbb{R}^{4\times 1}$ contains the physical quantities of the PDEs \eqref{eq:6}, \eqref{eq:7}
\begin{align}
&\bm{y}(\bm{x},t) = \begin{bmatrix}
p(\bm{x},t)&
\bm{i}\tran(\bm{x},t)
\end{bmatrix}\tran. \label{eq:15}
\end{align}
The temporal derivatives are captured by a temporal differential operator $\frac{\partial}{\partial t}\bm{D}$ including capacitance matrix $\bm{D}\in\mathbb{R}^{4\times 4}$, and the spatial differential operator $\Lp\in\mathbb{C}^{4\times 4}$ is composed of parameter matrix $\bm{A}\in\mathbb{R}^{4\times 4}$ and operator $\nabla\bm{B}$ containing spatial derivatives. Matrices $\bm{D}$, $\bm{A}$, and operator $\bm{\nabla}\bm{B}$ are given by  
\begin{align}
&\bm{D} = \begin{bmatrix}
\bm{0} &\bm{0}\\
1 & \bm{0}
\end{bmatrix}, 
&\bm{A} = \begin{bmatrix}
\bm{0} & -\bm{I} \\
0 & \bm{0}
\end{bmatrix}, 
&&\bm{\nabla}\bm{B} = \begin{bmatrix}
-D\, \mbox{grad} & \bm{0}\\
0 & -\mbox{div}
\end{bmatrix}, \label{eq:16}
\end{align}
with identity matrix $\bm{I}\in\mathbb{R}^{3\times 3}$. 
In accordance with the vector of variables $\bm{y}$ in \eqref{eq:15}, the source function $f_\mathrm{s}$ in \eqref{eq:6} is arranged into the vector $\bm{f}_\mathrm{e}\in\mathbb{C}^{4\times 1}$ in \eqref{eq:12} and initial condition \eqref{eq:11} into vector $\bm{y}_\mathrm{i}$ in \eqref{eq:14} as follows
\begin{align}
&\bm{f}_\mathrm{e}(\bm{x},t) = \begin{bmatrix}
0 & 0 & 0 &
f_\mathrm{s}(\bm{x},t)
\end{bmatrix}\tran,
&&\bm{y}_\mathrm{init}(\bm{x})= \begin{bmatrix}
p_\mathrm{init}(\bm{x}) & 0 & 0 & 0
\end{bmatrix}\tran.\label{eq:17_initSource}
\end{align}
The vector valued flow term in \eqref{eq:7} is included in vector $\bm{v}_\mathrm{flow}\in\mathbb{C}^{4\times 1}$ and is moved to the right hand side of  \eqref{eq:12}
\begin{align}
	\bm{v}_{\mathrm{flow}}(\bm{x},t) 
= \begin{bmatrix}
0&
0&
v(r)\, p(\bm{x},t)&
0
\end{bmatrix}\tran. \label{eq:17_flow}
\end{align}


\subsection{Laplace Transformation}
Before the proposed TFM can be derived, the mathematical time domain description \eqref{eq:12} is transformed into the continuous frequency domain. Application of the one-sided Laplace transform $\mathcal{L}\{\cdot\}$ to \eqref{eq:12} - \eqref{eq:14} yields an equivalent description in the frequency domain
\begin{align}
\left[s\bm{D} - \Lp \right]\bm{Y}(\bm{x},s) &= \bm{F}_\mathrm{e}(\bm{x},s) + \bm{V}_\mathrm{flow}(\x,s) + \bm{D}\bm{y}_\mathrm{init}(\bm{x}), 
&&\bm{x}\in V,\, s\in\mathbb{C},\label{eq:18}
\end{align}
where variables in the continuous frequency domain are denoted by upper-case letters and depend on the complex frequency variable $s$, i.e., $\mathcal{L}\{\bm{y}(\bm{x},t)\} = \bm{Y}(\bm{x},s)$. The temporal derivatives $\frac{\partial}{\partial t}$ in \eqref{eq:12} are replaced by multiplications with complex frequency variable $s\in\mathbb{C}$. The term $\bm{D}\bm{y}_\mathrm{init}$ on the right hand side of \eqref{eq:18} arises from the ICs \eqref{eq:11}. 
%


\section{Open Loop Transfer Function Model}
\label{sec:tfm1}
In this section, the vector $\bm{V}_\mathrm{flow}$ containing the term responsible for laminar flow is omitted, which simplifies \eqref{eq:18} to 
\begin{align}
\left[s\bm{D} - \Lp \right]\bm{Y}(\bm{x},s) &= \bm{F}_\mathrm{e}(\bm{x},s) + \bm{D}\bm{y}_\mathrm{init}(\bm{x}), 
&&\bm{x}\in V,\, s\in\mathbb{C}.\label{eq:19}
\end{align}
Omitting $\bm{V}_\mathrm{flow}$ reduces \eqref{eq:18} to a 3D diffusion problem. 
In the following, 
after some initial remarks regarding the modeling approach, its individual components are introduced. Then, the proposed approach is applied to \eqref{eq:19} with BCs \eqref{eq:9}, \eqref{eq:10} yielding a model for 3D diffusion in the cylinder. The derived model is formulated in terms of an SSD constituting the open loop system that forms the basis for the incorporation of $\bm{V}_\mathrm{flow}$ in Section~\ref{sec:tfm2}.

\subsection{Initial Remarks}
\vspace*{-1ex}
The applied modeling approach is based on the modal expansion of the vector PDE \eqref{eq:19} into an infinite set of bi-orthogonal eigenfunctions $\Kprim(\bm{x},\mu)\in\mathbb{C}^{4\times 1}$ and  $\Kadj(\bm{x},\mu)\in\mathbb{C}^{4\times 1}$ where the functions $\Kprim$ are the primal eigenfunctions and $\Kadj$ are their adjoints \cite{rabenstein:ijc:2017}. 
Furthermore, each eigenfunction $\Kprim$, $\Kadj$ is associated with an eigenvalue $s_\mu$, where the infinitely many eigenvalues define the discrete spectrum of the spatial differential operator $\Lp$ \cite{Curtain-Zwart:infdimsysth:1995,churchill:1972}. Although the exact form of the eigenvalues and eigenfunctions will be derived later in Section~\ref{subsec:model:openloop}, index $\mu \in \mathbb{Z}$ is already introduced here to count the eigenvalues and eigenfunctions.

The infinite number of eigenvalues and eigenfunctions is necessary to ensure convergence of the analytical solution. Nevertheless, for numerical evaluation in Section~\ref{sec:simul} only a finite number of eigenvalues can be considered. Therefore, the number of eigenvalues is truncated to $\mu = 0, \dots, Q - 1$ in the subsequent sections \cite[Chap.~4.8]{Schaefer2020}.

\subsection{Forward and Inverse Transformation}
\label{subsec:model:trafo}
\vspace*{-1ex}
The solution $\bm{Y}$ of \eqref{eq:19} in terms of an SSD is established by the application of a pair of transformations that are introduced in the following. Their main purpose is to transform the spatial derivatives in operator $\Lp$, where the transformation should have a similar effect as the Laplace transform does for the temporal derivatives. 
Based on well-known concepts from operator theory and functional analysis, the proposed spatial transformations constitute a forward and an inverse Sturm-Liouville transformation (SLT)\cite{churchill:1972}.

\subsubsection{Forward Transformation}

Forward transformation $\mathcal{T}\{\cdot\}$ performs an expansion of $\bm{Y}(\bm{x},s)$ into a set of $Q$ adjoint eigenfunctions $\tilde{\bm{K}} (\bm{x},\mu)$, where an individual expansion coefficient $\bar{Y}(\mu,s)$ can be defined in terms of a scalar product or an integral in $V$
\begin{align}
&\bar{Y}(\mu,s) = \linpr \bm{D}\bm{Y}(\bm{x},s), \tilde{\bm{K}} (\bm{x},\mu)\rinpr = \!\!\int_{V}\!\! \tilde{\bm{K}}\her (\bm{x},\mu) \bm{D}\bm{Y}(\bm{x},s)\,\dint{\bm{x}},
\label{eq:22}
\end{align}
where $(\cdot)\her$ denotes the conjugate-transpose.
Arranging the adjoint eigenfunctions $\tilde{\bm{K}} (\bm{x},\mu)$ into a matrix $\tilde{\Cs}(\bm{x})\in\mathbb{C}^{4 \times Q}$ and the expansion coefficients $\bar{Y}(\mu,s)$ into a vector $\bar{\bm{Y}}(s)\in\mathbb{C}^{Q\times 1}$,
\begin{align}
&\bar{\bm{Y}}(s) \!= \!\left[\bar{Y}(0,s), \dots,\,\bar{Y}(Q-1,s)  \right]\tran\!\!, 
&\!\!\tilde{\Cs}(\bm{x}) \!= \!\left[\Kadj(\bm{x},0), \dots,\,\Kadj(\bm{x},Q-1) \right],\label{eq:21}
\end{align} 
the forward transformation $\mathcal{T}\{\cdot\}$ is defined in terms of a vector-valued scalar product  
\begin{align}
&\mathcal{T}\left\lbrace \bm{Y}(\bm{x},s) \right\rbrace = \bar{\bm{Y}}(s) = \linpr \bm{D} \bm{Y}(\bm{x},s), \tilde{\Cs}(\bm{x})\rinpr
= \begin{bmatrix}
	\linpr \bm{D}\bm{Y}(\bm{x},s), \tilde{\bm{K}} (\bm{x},0)\rinpr\\
	\vdots\\
	\linpr \bm{D}\bm{Y}(\bm{x},s), \tilde{\bm{K}} (\bm{x},Q-1)\rinpr
\end{bmatrix}
, 
\label{eq:20}
\end{align}
wherein matrix $\tilde{\Cs}$ acts as transformation kernel.
%
In the following, variables that are transformed with \eqref{eq:20} are denoted by an overbar and by the variable's dependence on $\mu$ in the scalar case. 

\subsubsection{Differentiation Theorem}
The most important part of the forward transformation is the definition of a suitable differentiation theorem enabling the replacement of the spatial derivatives by the multiplication with a frequency domain variable. Therefore, to fit the forward transformation \eqref{eq:20}, a differentiation theorem for operator $\Lp$ in \eqref{eq:19} is defined as follows \cite{rabenstein:ijc:2017}
\begin{align}
\bm{\mathcal{T}}\left\lbrace \Lp\bm{Y}(\bm{x},s) \right\rbrace = \linpr \Lp \bm{Y}(\bm{x},s), \tilde{\Cs}(\bm{x})\rinpr = \As \bar{\bm{Y}}(s) + \bar{\bm{\Phi}}(s). \label{eq:23}
\end{align}
Diagonal matrix $\As = \mathrm{diag}\left(s_0,\,\dots,\, s_{Q-1}\right)\in\mathbb{C}^{Q\times Q}$ contains all $Q$ eigenvalues $s_\mu$, which act as spatial frequency variables for the transformation. Vector $\bar{\bm{\Phi}}\in\mathbb{C}^{Q\times 1}$ arises from the BCs -- analogous to the ICs in the Laplace transform \eqref{eq:18} -- and contains the transformed boundary values \cite{rabenstein:ijc:2017}. 
For the problem at hand, due to the homogeneous BCs \eqref{eq:9}, \eqref{eq:10}, $\bar{\bm{\Phi}}$ vanishes, i.e., $\bar{\bm{\Phi}} = \bm{0}$. Therefore, it is omitted in the subsequent derivations. 


\subsubsection{Application to the PDE}
Applying forward transformation \eqref{eq:20} to PDE \eqref{eq:19} and exploiting differentiation theorem \eqref{eq:23} yields a representation of \eqref{eq:19} in a spatio-temporal transform domain
\begin{align}
\left[s\bm{D} - \Lp\right]\bm{Y}(\bm{x},s) &= \bm{F}_\mathrm{e}(\bm{x},s) + \bm{D}\bm{y}_\mathrm{init}&&\big\vert \langle\, \cdot\,, \tilde{\Cs}(\bm{x})\rangle \nonumber\\
s\,\langle\bm{D} \bm{Y}(\bm{x},s), \tilde{\Cs}(\bm{x})\rangle - \langle\Lp\bm{Y}(\bm{x},s), \tilde{\Cs}(\bm{x})\rangle &= \langle\bm{F}_\mathrm{e}(\bm{x},s), \tilde{\Cs}(\bm{x})\rangle + 
 \langle\bm{D}\bm{y}_\mathrm{init}(\bm{x}), \tilde{\Cs}(\bm{x})\rangle \nonumber\\
s\bar{\bm{Y}}(s) - \As\bar{\bm{Y}}(s) &= \bar{\bm{F}}_\mathrm{e}(s) + \bar{\bm{y}}_\mathrm{init}. \label{eq:24}
\end{align}
Here, vectors $\bar{\bm{F}}_\mathrm{e}\in\mathbb{C}^{Q\times 1}$ and $\bar{\bm{y}}_\mathrm{init}\in\mathbb{C}^{Q\times 1}$ are the transform domain representations of the external sources $\bm{F}_\mathrm{e}$ and the initial conditions $\bm{y}_\mathrm{init}(\bm{x})$.

\subsubsection{Inverse Transformation}

For forward transformation \eqref{eq:20}, an inverse transformation $\mathcal{T}^{-1}\{\cdot\}$ is defined. The inverse transformation exploits the discrete nature of the spectrum of operator $\Lp$, which allows its formulation in terms of a generalized Fourier series \cite{Curtain-Zwart:infdimsysth:1995, churchill:1972}
\begin{align}
\mathcal{T}^{-1}\!\!\left\lbrace \bm{\bar{Y}}(s) \right\rbrace &= \bm{Y}(\bm{x},s) = \sum_{\mu = 0}^{Q-1}\frac{1}{N_\mu}\bar{Y}(\mu,s)\,\Kprim(\bm{x},\mu). \label{eq:25}
\end{align}
To fit the formulation of forward transformation \eqref{eq:20}, the sum in \eqref{eq:25} is reformulated in terms of a matrix-valued transformation kernel $\Cs\in\mathbb{C}^{4\times Q}$ containing primal eigenfunctions $\Kprim$ and scaling factor $N_\mu$
\begin{align}
&\mathcal{T}^{-1}\!\!\left\lbrace \bm{\bar{Y}}(s) \right\rbrace = \bm{Y}(\bm{x},s) = \Cs(\bm{x}) \bar{\bm{Y}}(s), 
&\Cs(\bm{x}) = \left[\frac{1}{N_0}\Kprim(\bm{x},0), \dots,\, 
\frac{1}{N_{Q-1}}\Kprim(\bm{x},Q-1)
\right]. \label{eq:26}
\end{align}
Scaling factor $N_\mu$ originates from the bi-orthogonality of eigenfunctions $\Kprim$ and $\Kadj$ and is required for the formulation of an inverse transformation \cite[Chap.~4.7]{Schaefer2020}
\begin{align}
N_\mu = \langle \bm{D}\Kprim(\bm{x},\mu), \Kadj(\bm{x},\mu)\rangle. \label{eq:27}
\end{align}

\subsection{Eigenfunctions and Eigenvalues}
\label{subsec:model:ev}
\vspace*{-1ex}
To obtain an analytical solution, eigenfunctions $\Kprim$, $\Kadj$, eigenvalues $s_\mu$, and scaling factor $N_\mu$ have to be derived. These variables are derived based on the underlying physical system exploiting specific properties of Sturm-Liouville theory. 
These properties are not presented in detail here but can be found in the relevant literature, e.g., \cite{churchill:1972,Curtain-Zwart:infdimsysth:1995,Ering:54}. 
\subsubsection{Eigenfunctions} Eigenfunctions $\Kprim$ and $\Kadj$ are derived by evaluation of the corresponding eigenvalue problems, which are well established for PDEs in the form of \eqref{eq:19}. The eigenvalue problem for the primal eigenfunctions $\Kprim$ is \cite{rabenstein:ijc:2017} 
\begin{align}
\Lp\Kprim(\bm{x},\mu) &= s_\mu\bm{D}\Kprim(\bm{x},\mu),\label{eq:28}
\end{align}
where the eigenfunctions have to fulfill homogeneous BCs on $\partial V_\mathrm{r}$, $\partial V_\mathrm{z}$ that are closely related to BCs \eqref{eq:9}, \eqref{eq:10}. The evaluation of \eqref{eq:28} strongly depends on the exact form of $\Lp$ which, in the considered case, consists of gradient and divergence operators \eqref{eq:16}. In this particular case, classical separation of variables can be applied and the resulting solution is well known in the context of heat transfer \cite{carslaw:1946}, and has been recently used in the context of MC in \cite{Zoofaghari:ieee:2019}. 
Similar to \eqref{eq:28}, an eigenvalue problem for the adjoint eigenfunctions $\Kadj$ can be established, but is omitted here for brevity \cite{rabenstein:ijc:2017}.
The resulting eigenfunctions can be organized in vector form as follows
\begin{align}
&\Kprim(\bm{x},\mu) = \begin{bmatrix}
J_n(k_{n,m}r)\expE{\jcomp n \varphi}\sin(\lambda_\nu z)\\
- D\, k_{n,m}\,J_n'(k_{n,m}r)\expE{\jcomp n \varphi}\sin(\lambda_\nu z)\\
- D\, \frac{\jcomp n}{r}\,J_n(k_{n,m}r)\expE{\jcomp n \varphi}\sin(\lambda_\nu z)\\
-D\,\lambda_\nu J_n(k_{n,m}r)\expE{\jcomp n \varphi}\cos(\lambda_\nu z)
\end{bmatrix},
&\Kadj(\bm{x},\mu) = \begin{bmatrix}
k_{n,m}\,J_n'(k_{n,m}r)\expE{\jcomp n \varphi}\sin(\lambda_\nu z)\\
\frac{\jcomp n}{r}\,J_n(k_{n,m}r)\expE{\jcomp n \varphi}\sin(\lambda_\nu z)\\
\lambda_\nu J_n(k_{n,m}r)\expE{\jcomp n \varphi}\cos(\lambda_\nu z)\\
J_n(k_{n,m}r)\expE{\jcomp n \varphi}\sin(\lambda_\nu z)
\end{bmatrix}, \label{eq:29}
\end{align}
where $J_n(x)$ denotes the $n$th order Bessel function of the first kind, $J'(x)= \frac{\partial}{\partial x}J(x)$ is the corresponding derivative, and $\jcomp = \sqrt{-1}$. The values $k_{n,m}$ and $\lambda_\nu$ with indices $m\in\mathbb{Z}$ and $\nu\in\mathbb{Z}$ are related to the eigenvalues $s_\mu$ and will be provided in the following. 
%
%

\subsubsection{Eigenvalues} Eigenvalues $s_\mu$ are derived from homogeneous BCs that have to be fulfilled by $\Kprim$ and $\Kadj$. The corresponding derivation is omitted for brevity, but a detailed description can be found in, e.g., \cite{rabenstein:ijc:2017}. 
The BC \eqref{eq:9} on $\partial V_\mathrm{z}$ is a condition for the first entry of $\bm{y}$ in \eqref{eq:15}, and therefore, the first entry $K_1$ of $\Kprim$ in \eqref{eq:29} has to fulfill a homogeneous BC on $\partial V_\mathrm{z}$ yielding the condition
\begin{align}
&K_1(\bm{x},\mu)\big\vert_{z = Z_0} \overset{!}{=} 0 &\rightarrow && \sin\left(\lambda_\nu Z_0\right) \overset{!}{=} 0, \label{eq:31}
\end{align}
where relation \eqref{eq:31} is fulfilled by wave-numbers $\lambda_\nu$ of the form $\lambda_\nu = \nu \frac{\pi}{Z_0}$. 
The BC \eqref{eq:10} on $\partial V_\mathrm{r}$ is a condition for the second entry of $\bm{y}$ in \eqref{eq:15}, and therefore, the second entry $K_2$ of $\Kprim$ in \eqref{eq:29} has to fulfill homogeneous BCs on $\partial V_\mathrm{r}$ yielding the condition
\begin{align}
&K_2(\bm{x},\mu)\big\vert_{r = R_0} \overset{!}{=} 0  &\rightarrow &&k_{n,m}\,J_n'(k_{n,m}R_0) \overset{!}{=} 0,
\label{eq:30}
\end{align}
where $k_{n,m}$ is the $m$-th real-valued root of $J_n'(k_{n,m}R_0)$.
Finally, the eigenvalues $s_\mu$ of the system are defined in terms of the roots $k_{n,m}$ and wave-numbers $\lambda_\nu$
\begin{align}
s_\mu = s_{n,m,\nu} = -D\left(k_{n,m}^2 + \lambda_\nu^2\right). \label{eq:32}
\end{align}
This equation reveals the relevance of index $\mu$, i.e., its purpose to count the individual eigenvalues $s_\mu$. $s_\mu$ depends on the $k_{n,m}$, i.e., on order $n$ and root index $m$ in \eqref{eq:30}, and on index $\nu$ of wave-numbers $\lambda_\nu$ in \eqref{eq:31}. Particularly, $\mu$ represents an index tupel, i.e., $(n, m, \nu) \to \mu$. 
%

\subsubsection{Scaling Factor}
Scaling factor $N_\mu$ is defined in \eqref{eq:27} and due to the geometrical separability of the cylindrical system, it is separated into three individual components 
\begin{align}
N_\mu = \langle \bm{D}\Kprim(\bm{x},\mu), \Kadj(\bm{x},\mu)\rangle = N_r(\mu)\,N_\varphi(\mu)\,N_z(\mu).
\label{eq:33}
\end{align}
Evaluating the scalar product in \eqref{eq:33} in terms of an integral, see \eqref{eq:22}, 
the individual components of $N_\mu$ can be obtained as follows
\begin{align}
N_r(\mu) &= 
\int_{0}^{R_0}J_n^2(k_{n,m}r)r\dint{r} &&= 
\begin{cases}
\frac{R_0}{2} & k_{n,m} = 0\\
\frac{R_0}{2}\frac{1}{k_{n,m}^2}\left(k_{n,m}^2 - \frac{n^2}{R_0^2}\right)J_n^2(k_{n,m}R_0) & k_{n,m} \neq 0
\end{cases}\label{eq:34}\\
%
N_\varphi(\mu) &= 
\int_{-\pi}^{\pi}\expE{\jcomp(n-n)\varphi}\dint{\varphi} &&= 2\pi \label{eq:35}\\
%
%
N_z(\mu) &= 
\int_{0}^{Z_0}\sin^2\lambda_\nu z\dint{z} &&= 
\begin{cases}
0 & \lambda_\nu = 0\\
\frac{Z_0}{2} & \lambda_\nu \neq 0
\end{cases}\label{eq:36}
\end{align}


\subsection{Open Loop Transfer Function Model}
\label{subsec:model:openloop}
\vspace*{-1ex}
Based on the preceding sections, the open loop TFM as the solution of the 3D diffusion process in \eqref{eq:19} can be obtained. The transform domain representation of PDE \eqref{eq:19} in \eqref{eq:24} serves as a \textit{state equation} with state vector $\bar{\bm{Y}}$. 
The inverse transformation \eqref{eq:26} acts as \textit{output equation}. Together, state equation \eqref{eq:24} and output equation \eqref{eq:26} constitute an SSD in the $s$-domain
\begin{align}
s\bar{\bm{Y}}(s) &= \As\bar{\bm{Y}}(s) + \bar{\bm{F}}_\mathrm{e}(s) + \bar{\bm{y}}_\mathrm{init}, \label{eq:37}\\
\bm{Y}(\bm{x},s) &= \Cs(\bm{x})\bar{\bm{Y}}(s).\label{eq:38}
\end{align}
The vector valued solution of PDE \eqref{eq:19}, i.e., vector $\bm{Y}$ in \eqref{eq:38}, contains different physical quantities. For the subsequent derivations and analysis in Sections~\ref{sec:simul} and \ref{sec:analysis}, the particle concentration is of interest. Therefore, output equation \eqref{eq:38} is reduced to deliver a solution for the particle concentration in the cylinder by restricting $\Cs$ in \eqref{eq:26} to its first row, i.e., vector $\bm{c}_1\tran\in\mathbb{C}^{1\times Q}$ 
\begin{align}
P(\bm{x},s) &= \bm{c}_1\tran(\bm{x})\bar{\bm{Y}}(s), \label{eq:39}\\
\bm{c}_1\tran(\bm{x}) &= 
\left[\frac{1}{N_0}K_1(\bm{x},0),\, \dots,\, \frac{1}{N_{Q-1}}K_1(\bm{x},Q-1)\right], 
\label{eq:39_c1}
%
\end{align}
where $K_1(\bm{x},\mu) = J_n(k_{n,m}r)\,\expE{\jcomp n \varphi}\sin\left(\lambda_\nu z\right)$ is the first entry of $\Kprim$ in \eqref{eq:29}.
We note that the representation of the solution by \eqref{eq:37} and \eqref{eq:39} has to be transformed into the continuous or discrete-time domain for analysis and numerical evaluation. 

\section{Closed Loop Transfer Function Model}
\label{sec:tfm2}
The previously derived open loop model \eqref{eq:37}, \eqref{eq:39} constitutes a solution for 3D diffusion in the cylinder. In this section, the previously excluded flow term $\bm{V}_\mathrm{flow}$ is reincorporated into the model to obtain a solution for the considered advection-diffusion process with laminar flow. To provide a clear starting point, \eqref{eq:19} is rewritten with the flow term included
\begin{align}
	\left[s\bm{D} - \Lp\right]\bm{Y}(\bm{x},s) &= \bm{F}_\mathrm{e}(\bm{x},s) +
\bm{D}\bm{y}_\mathrm{init}(\bm{x}) + 
\bm{V}_\mathrm{flow}(\bm{x},s). \label{eq:40}
\end{align}
Applying forward transformation \eqref{eq:20} to \eqref{eq:40} leads to a representation in the spatio-temporal transform domain 
\begin{align}
	s\bar{\bm{Y}}(s) &= \As\bar{\bm{Y}}(s) + \bar{\bm{F}}_\mathrm{e}(s) + 
\bar{\bm{V}}_\mathrm{flow}(s) + 
\bar{\bm{y}}_\mathrm{init}, \label{eq:41}
\end{align}
where vector $\bar{\bm{V}}_\mathrm{flow}\in\mathbb{C}^{Q\times 1}$ denotes the transform domain representation of $\bm{V}_\mathrm{flow}$, i.e.,
\begin{align}
\bar{\bm{V}}_\mathrm{flow}(s) &= \langle \bm{V}_\mathrm{flow}(\bm{x},s), \tilde{\Cs}(\bm{x})\rangle = \int_{V}\tilde{\Cs}\her(\bm{x}) \bm{V}_\mathrm{flow}(\bm{x},s)\dint{\bm{x}}. \label{eq:42}
\end{align}
A direct evaluation of \eqref{eq:42} is not possible in closed form as $\bm{V}_\mathrm{flow}$ itself depends on the particle concentration $P$, see \eqref{eq:17_flow}. 
%
In this paper, our objective is to establish an analytical solution for the considered advection-diffusion problem. Therefore, we express $\bar{\bm{V}}_\mathrm{flow}$ in \eqref{eq:42} in terms of the system states $\bar{\bm{Y}}$ and suitable feedback matrices. Finally, the derived expressions will introduce a feedback system which extends the open loop state equation \eqref{eq:37} to account for laminar flow.

%

\subsection{Decomposition of the Flow Profile}
\vspace*{-1ex}
Due to the structure of the flow profile $v(r)$ in \eqref{eq:5}, it can be decomposed into a uniform flow term and a parabolic term. Starting with \eqref{eq:42} and exploiting the structure of $\tilde{\Cs}$ in \eqref{eq:21} and $\bm{V}_\mathrm{flow}$ in \eqref{eq:17_flow} leads to a representation with separate uniform flow and parabolic terms
\begin{align}
\bar{\bm{V}}_\mathrm{flow}(s) &= \int_{V}\tilde{\bm{c}}_3^*(\bm{x})v(r)P(\bm{x},s)\dint{\bm{x}} 
=  \underbrace{v_0\int_{V}\tilde{\bm{c}}_3^*(\bm{x})P(\bm{x},s)\dint{\bm{x}}}_{=\bar{\bm{V}}_\mathrm{uni}(s)}
- 
\underbrace{
\frac{v_0}{R_0^2}\int_{V}\tilde{\bm{c}}_3^*(\bm{x})P(\bm{x},s) r^2\dint{\bm{x}}
}_{= \bar{\bm{V}}_\mathrm{par}(s)}.
\label{eq:43} 
%
\end{align}
Here, vector $\tilde{\bm{c}}_3\tran\in\mathbb{C}^{1\times Q}$ is the third row of $\tilde{\Cs}$ in \eqref{eq:21} 
\begin{align}
\tilde{\bm{c}}_3\tran(\bm{x}) = 
\left[\tilde{K}_3(\bm{x},0), \,\dots,\, \tilde{K}_3(\bm{x},Q-1)\right]
, \label{eq:44}
\end{align}
where $\tilde{K}_3(\bm{x},\mu) = \lambda_\nu J_n(k_{n,m}r)\expE{\jcomp n \varphi}\cos\left(\lambda_\nu z\right)$ denotes the third entry of $\Kadj$ in \eqref{eq:29}.
In the following, the terms $\bar{\bm{V}}_\mathrm{uni}$ and $\bar{\bm{V}}_\mathrm{par}$ are considered separately.

\subsection{Uniform Flow}
\vspace*{-1ex}
In the following, the uniform flow term $\bar{\bm{V}}_\mathrm{uni}$ is reformulated in terms of a feedback matrix and the open loop system states $\bar{\bm{Y}}$. The starting point is the representation of $\bar{\bm{V}}_\mathrm{uni}$ in terms of an integral in \eqref{eq:43}
\begin{align}
\bar{\bm{V}}_\mathrm{uni}(s) = v_0\langle P(\bm{x},s), \tilde{\bm{c}}_3\tran(\bm{x})\rangle = 
v_0\int_{V}\tilde{\bm{c}}_3^*(\bm{x})P(\bm{x},s)\dint{\bm{x}}. \label{eq:45}
\end{align}
Now, particle concentration $P$ is expressed in \eqref{eq:39} in terms of the system states $\bar{\bm{Y}}$ and vector $\bm{c}_1\tran$. Inserting this representation into \eqref{eq:45} leads to
\begin{align}
\bar{\bm{V}}_\mathrm{uni}(s) = 
v_0\int_{V}\tilde{\bm{c}}_3^*(\bm{x}) \bm{c}_1\tran(\bm{x})\dint{\bm{x}} \,\bar{\bm{Y}}(s) = 
v_0\Ks_\mathrm{uni}\bar{\bm{Y}}(s), \label{eq:46}
\end{align}
where feedback matrix $\Ks_\mathrm{uni}\in\mathbb{C}^{Q\times Q}$ is defined as
\begin{align}
	\Ks_\mathrm{uni} = \linpr \bm{c}_1\tran(\bm{x}), \tilde{\bm{c}}_3\her(\bm{x})\rinpr = 
	\int_{V}\tilde{\bm{c}}_3^*(\bm{x}) \bm{c}_1\tran(\bm{x})\dint{\bm{x}}. \label{eq:46a}
\end{align}
%
Exploiting the similar structure of $\bm{c}_1$ and $\tilde{\bm{c}}_3$, \eqref{eq:46} can be simplified exploiting, e.g., integral and orthogonality theorems for the involved Bessel functions, and a closed-form expression for \eqref{eq:46a} can be obtained. 
Furthermore, we note that matrix $\Ks_\mathrm{uni}$ is independent of the flow velocity $v_0$, but depends on the geometry of the cylinder. 
Thus, in practice, matrix $\Ks_\mathrm{uni}$ has to be calculated only once for a given cylinder geometry and can subsequently be scaled depending on the flow velocity $v_0$.

\subsection{Parabolic Flow Profile}
\vspace*{-1ex}
Analogous to the uniform flow term in \eqref{eq:45}, \eqref{eq:46}, the parabolic flow term $\bar{\bm{V}}_\mathrm{par}$ is reformulated. Starting point is its representation in terms of an integral in \eqref{eq:43}
\begin{align}
\bar{\bm{V}}_\mathrm{par}(s) = - \frac{v_0}{R_0^2}\langle P(\bm{x},s)r^2, \tilde{\bm{c}}_3\tran(\bm{x})\rangle = 
-\frac{v_0}{R_0^2}\int_{V}\tilde{\bm{c}}_3^*(\bm{x})P(\bm{x},s)r^2\dint{\bm{x}}. \label{eq:47}
\end{align}
Similar to \eqref{eq:46}, the particle concentration is expressed by \eqref{eq:39}, which is inserted into \eqref{eq:47}. This leads to a representation of $\bar{\bm{V}}_\mathrm{par}$ in terms of matrix $\Ks_\mathrm{par}\in\mathbb{C}^{Q\times Q}$ and system states $\bar{\bm{Y}}$,
\begin{align}
\bar{\bm{V}}_\mathrm{par}(s) = 
- \frac{v_0}{R_0^2}\int_{V}\tilde{\bm{c}}_3^*(\bm{x}) \bm{c}_1\tran(\bm{x})r^2\dint{\bm{x}} \,\bar{\bm{Y}}(s) = 
- \frac{v_0}{R_0^2}\Ks_\mathrm{par}\bar{\bm{Y}}(s), \label{eq:48}
\end{align}
where
\begin{align}
	\Ks_\mathrm{par} = \linpr \bm{c}_1\tran(\bm{x})\,r^2, \tilde{\bm{c}}_3\her(\bm{x})\rinpr = 
	\int_{V}\tilde{\bm{c}}_3^*(\bm{x}) \bm{c}_1\tran(\bm{x})\,r^2 \dint{\bm{x}}. \label{eq:48a}
\end{align}
Similar to $\Ks_\mathrm{uni}$, matrix $\Ks_\mathrm{par}$ can also be pre-calculated and depends only on the geometry of the cylinder. However, in contrast to $\Ks_\mathrm{uni}$, matrix $\Ks_\mathrm{par}$ can not be obtained in closed form except for Bessel functions of order $n = 0$.

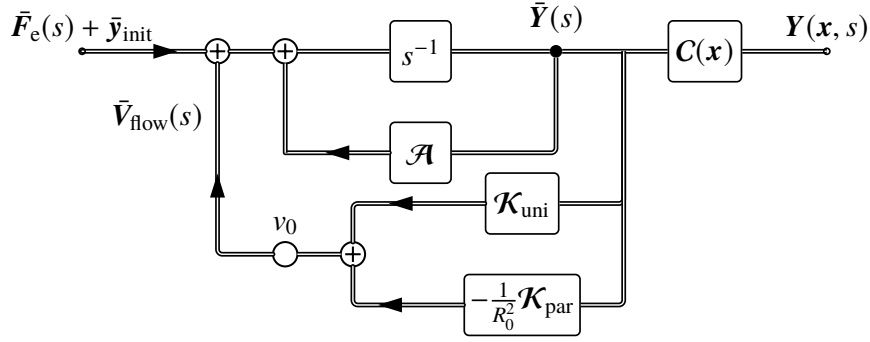
\begin{figure}[t]
	\centering
	\scalebox{0.9}{
	\begin{tikzpicture}[auto, scale = 1, every node/.style={scale=0.9}, font=\large, node distance=2.5cm,>=latex', 
	rounded corners=2pt]
	\node[draw, dspsquare, minimum width=1cm, minimum height=1cm](s)at (8, 4) {
		\parbox{1cm}{
			\centering $s^{-1}$
	}};
	\node[draw, dspsquare, minimum width=1cm, minimum height=1cm](a)at (8, 2.5) {
		\parbox{1cm}{
			\centering $\bm{\mathcal{A}}$
	}};
	
	\node[draw, dspsquare, minimum width=1cm, minimum height=1cm](klam)at (9.5, 0.25) {
		\parbox{1.9cm}{
			\centering $-\frac{1}{R_0^2}\Ks_\mathrm{par}$
	}};

	\node[draw, dspsquare, minimum width=1cm, minimum height=1cm](kuni)at (9.5, 1.75) {
		\parbox{1.2cm}{
			\centering $\Ks_\mathrm{uni}$
	}};
	
	\node[draw,dspadder](add) at (6,4){};
	\node[draw,dspadder](add2) at (5,4){};
	\node[draw,dspadder](add3) at (7,1){};
	\node[draw,dspmultiplier](mu) at (6,1){$v_0$};
	\node[draw,circle,fill,inner sep = 0pt, minimum size=1ex](p) at (10,4){};
	\node[draw, dspsquare, minimum width=1cm, minimum height=1cm](c)at (12.2, 4) {
		\parbox{1.2cm}{
			\centering $\bm{\mathcal{C}}(\bm{x})$
	}};
	\draw[dspflow,double] (3,4) node[dspnodeopen]{$\bar{\bm{F}}_\mathrm{e}(s) + \bar{\bm{y}}_\mathrm{init}$} to (add2) -- (add);
	\draw[dspline,double] (add) to (s) to (p);
	\draw[dspline,double] (p) to (10,2.5) to (a);
	\draw[dspflow,double] (a) to (6,2.5); 
	\draw[dspline,double] (6,2.5) to (add);
	\draw[dspline,double] (p) to (c);
	\draw[dspline,double] (c) to (14,4) node[dspnodeopen]{$\bm{Y}(\bm{x},s)$}; 
	
	\draw[dspline,double] (11,4) |- (kuni);
	\draw[dspline,double] (11,4) |- (klam);
	\draw[dspflow,double] (kuni) -| (add3);	
	\draw[dspflow,double] (klam) -| (add3);	
	\draw[dspline,double] (add3) -- (mu);
	\draw[dspflow,double] (mu) -| node[midway,left,xshift = -0.5ex, yshift = 12ex]{$\bar{\bm{V}}_{\mathrm{flow}}(s)$}(add2);
	
	\node at (10,4.5) {
		\parbox{3cm}{
			\centering $\bar{\bm{Y}}(s)$
	}};
	\end{tikzpicture}
	}
	\vspace*{-2ex}
	\caption{\small Closed loop state space description of the advection-diffusion system in \eqref{eq:6}, \eqref{eq:7} based on the derived TFM with state equation \eqref{eq:49} and output equation \eqref{eq:38}. }
	\label{fig:3}
	\vspace*{-3ex}
\end{figure}

\subsection{Closed Loop Transfer Function Model}
\vspace*{-1ex}
Inserting $\bar{\bm{V}}_\mathrm{uni}$ and $\bar{\bm{V}}_\mathrm{par}$ into \eqref{eq:43} and subsequently $\bar{\bm{V}}_\mathrm{flow}$ into \eqref{eq:41} leads to the closed loop state equation with modified velocity dependent state matrix $\As_\mathrm{c}$ 
\begin{align}
s\bar{\bm{Y}}(s) &= \As_\mathrm{c}(v_0)\bar{\bm{Y}}(s) + \bar{\bm{F}}_\mathrm{e}(s) + \bar{\bm{y}}_\mathrm{init},
&&\As_\mathrm{c}(v_0) = \As + v_0 \left(\Ks_\mathrm{uni} - \frac{1}{R_0^2}\Ks_\mathrm{par}\right), \label{eq:49}
\end{align}
while output equation \eqref{eq:38} remains unchanged. 
Together, state equation \eqref{eq:49} and output equation \eqref{eq:38} constitute the closed loop TFM which is a solution to the considered advection-diffusion problem in the $s$-domain. In Fig.~\ref{fig:3}, the complete SSD is illustrated. 
The figure clearly shows that matrices $\Ks_\mathrm{uni}$ and $\Ks_\mathrm{par}$ constitute feedback matrices in the proposed SSD model. 

Structures as shown in Fig.~\ref{fig:3} are well known in control theory for the design of parametric feedback control systems \cite{Deutscher:book:2012}. 
In the considered scenario, the feedback structure serves a different purpose but its principle influence on the overall system behavior is similar. 
In particular, the open loop system is characterized by matrix $\As$ containing the eigenvalues of the 3D diffusion process in the absence of flow (see Section~\ref{sec:tfm1}). 
The flow term is incorporated via the two feedback matrices which act on matrix $\As$. Particularly, feedback matrices $\Ks_\mathrm{uni}$ and $\Ks_\mathrm{par}$ shift the eigenvalues in $\As$ which results in the new matrix $\As_\mathrm{c}$. Therefore, the impact of flow has been reduced to a modification of the eigenvalues in $\As$ resulting in the new state matrix $\As_\mathrm{c}$ which fully captures the behavior of the advection-diffusion system including laminar flow.

\subsection{Initial Conditions and External Sources}
\label{subsec:ext}
\vspace*{-1ex}
The closed loop state equation \eqref{eq:49} of the advection-diffusion process contains -- yet unspecified -- source terms, i.e., functions $\bar{\bm{F}}_\mathrm{e}$ and $\bar{\bm{y}}_\mathrm{init}$ of the external sources and initial conditions, respectively. 
Both functions and their properties are discussed in the following. 

\subsubsection{Initial Conditions}

Via initial conditions $\bm{y}_\mathrm{init}$ in \eqref{eq:14}, an initial distribution of particles $p_\mathrm{init}$ can be defined in the cylinder volume $V$.
To obtain the transform domain representation $\bar{\bm{y}}_\mathrm{init}$ in \eqref{eq:49}, the initial conditions $\bm{y}_\mathrm{init}$ in the space domain have to be transformed as in \eqref{eq:24}
\begin{align}
\bar{\bm{y}}_\mathrm{init} = \langle \bm{D}\bm{y}_\mathrm{init}(\bm{x}), \tilde{\Cs}(\bm{x})\rangle = 
\int_{V}\tilde{\Cs}\her(\bm{x})\bm{D}\bm{y}_\mathrm{init}(\bm{x})\dint{\bm{x}}. \label{eq:50}
\end{align}
The integral can be simplified exploiting the structure of  $\bm{D}$ in \eqref{eq:16} and $\bm{y}_\mathrm{init}$ in \eqref{eq:17_initSource} as follows
\begin{align}
\bar{\bm{y}}_\mathrm{init} &= \int_{V} \tilde{\bm{c}}_4^*(\bm{x})\,p_\mathrm{init}(\bm{x})\dint{\bm{x}} = \langle p_\mathrm{init}(\bm{x}), \tilde{\bm{c}}\tran_4(\bm{x}) \rangle,\label{eq:51}\\
\tilde{\bm{c}}_4\tran(\bm{x}) &= 
\left[\tilde{K}_4(\bm{x},0), \,\dots,\, \tilde{K}_4(\bm{x},Q-1)\right] 
, \label{eq:51_c4}
%
\end{align}
where $\tilde{\bm{c}}_4\tran\in\mathbb{C}^{1\times Q}$ is the fourth row of matrix $\tilde{\Cs}$ in \eqref{eq:21} and $\tilde{K}_4(\bm{x},\mu) = J_n(k_{n,m}r)\expE{\jcomp n \varphi}\sin\left(\lambda_\nu z\right)$ is the fourth entry of $\Kadj$ in \eqref{eq:29}. 
%

\subsubsection{External Sources} Via function $\bm{F}_\mathrm{e}$ in \eqref{eq:18}, i.e., via function $f_\mathrm{s}$ in \eqref{eq:6}, it is possible to model the spatial and temporal distributions of the injected particles. We assume that the corresponding function $f_\mathrm{s}$ in \eqref{eq:6} in the continuous-time domain is separable, i.e.,  
\begin{align}
f_\mathrm{s}(\bm{x},t) = f_\mathrm{t}(t)\cdot f_\mathrm{x}(\bm{x}), \label{eq:52}
\end{align}
where $f_\mathrm{t}$ models the temporal pulse shaping of an injection and $f_\mathrm{x}$ models the spatial distribution. 
Similar to the initial conditions \eqref{eq:50}, the transform domain representation $\bar{\bm{F}}_\mathrm{e}$ is obtained by the transformation in \eqref{eq:24}
\begin{align}
\bar{\bm{F}}_\mathrm{e}(s) = \langle \bm{F}_\mathrm{e}(\bm{x},s), \tilde{\Cs}(\bm{x})\rangle = 
\int_{V}\tilde{\Cs}\her(\bm{x})\bm{F}_\mathrm{e}(\bm{x},s)\dint{\bm{x}}, \label{eq:53}
\end{align}
which can be simplified by exploiting the structure of \eqref{eq:17_initSource} and the separability assumed in \eqref{eq:52}
\begin{align}
&\bar{\bm{F}}_\mathrm{e}(s) = \int_{V} \tilde{\bm{c}}_4^*(\bm{x})\,F_\mathrm{s}(\bm{x},s)\dint{\bm{x}} = 
F_\mathrm{t}(s)\int_{V}\tilde{\bm{c}}_4^*(\bm{x})\,F_\mathrm{x}(\bm{x})\dint{\bm{x}}. \label{eq:54}
\end{align}

\subsection{Relation to Green's Function}
\vspace*{-1ex}
As mentioned in Section~\ref{sec:tfm1}, the proposed modeling approach is based on the modal expansion of an IBVP. Modeling MC channels via modal expansions is well established for regular shapes such as cylinders, see, e.g., \cite{Zoofaghari:ieee:2019}, and often the channel is finally modeled in terms of a CGF. 
Although the presented approach differs from classical ones, especially due to its ability to incorporate complex flow profiles, the obtained solution can be related to a representation in terms of a CGF. 
To this end, first, the continuous-time equivalents of state equation \eqref{eq:49} and output equation \eqref{eq:38} are derived by application of an inverse Laplace transform
\begin{align}
s\bar{\bm{Y}}(s) &= \As_\mathrm{c}(v_0)\bar{\bm{Y}}(s) + \bar{\bm{y}}_\mathrm{init},
&P(\bm{x},s) &= \bm{c}_1\tran(\bm{x})\bar{\bm{Y}}(s), \label{eq:55}\\
&\TransformVert\,\mathcal{L}^{-1}\{\cdot\} & &\TransformVert\,\mathcal{L}^{-1}\{\cdot\} \nonumber\\
\bar{\bm{y}}(t) &= \expE{\As_\mathrm{c}(v_0)\,t}\bar{\bm{y}}_\mathrm{init}, 
&p(\bm{x},t) &= \bm{c}_1\tran(\bm{x})\bar{\bm{y}}(t).\label{eq:56}
\end{align}
Function $\bar{\bm{F}}_\mathrm{e}$ is omitted for the following considerations as the CGF is usually only derived with initial conditions. Inserting $\bar{\bm{y}}(t)$ and $\bar{\bm{y}}_\mathrm{init}$ from \eqref{eq:50} into  output equation \eqref{eq:56} leads to 
\begin{align}
p(\bm{x},t) = \bm{c}_1\tran(\bm{x}) \expE{\As_\mathrm{c}(v_0)\,t}\langle p_\mathrm{init}(\bm{x}), \tilde{\bm{c}}\tran_4(\bm{x}) \rangle. \label{eq:57}
\end{align}
Exploiting the integral formulation of \eqref{eq:50} and rearranging \eqref{eq:57} yields a representation of the concentration in terms of a Green's function, i.e., the CGF of the advection-diffusion problem
\begin{align}
&p(\bm{x},t) = \int_{V} g(t, \bm{x}\vert \bm{\xi}) p_\mathrm{init}(\bm{\xi})\dint{\bm{\xi}}, 
&g(t, \bm{x}\vert \bm{\xi}) = \bm{c}_1\tran(\bm{x}) \,\expE{\As_\mathrm{c}(v_0)\,t}\tilde{\bm{c}}^*_4(\bm{\xi}),
\label{eq:58}
\end{align}
with spatial integration variable $\bm{\xi}$.
The Green's function, $g$, in \eqref{eq:58} is expressed in terms of the eigenfunctions $\bm{c}_1$ in \eqref{eq:39_c1} and  $\tilde{\bm{c}}_4$ in \eqref{eq:51_c4}, and modified state matrix $\As_\mathrm{c}$, which includes the impact of laminar flow. 
The influence of source function $\bar{\bm{f}}_\mathrm{e}$ can be incorporated by convolution with \eqref{eq:58}.


\subsection{Interpretation in Terms of Transfer Functions}
\vspace*{-1ex}
The proposed model is based on transfer functions. To make this fact more explicit, the representation in terms of an SSD is reformulated by inserting state equation \eqref{eq:49} into output equation \eqref{eq:39} and solving for the concentration
\begin{align}
	&P(\bm{x},s) = \bm{c}_1\tran(\bm{x})\bar{\bm{H}}(s,D,v_0)\left[\bar{\bm{F}}_\mathrm{e}(s) + \bar{\bm{y}}_\mathrm{init} \right], 
	&\bar{\bm{H}}(s,D,v_0) = \left(s\bm{I} - \As_\mathrm{c}(v_0,D)\right)^{-1}, \label{eq:60a}
\end{align}
where $\bar{\bm{H}}$ denotes the transfer function. For clarity, the dependence of $\As_\mathrm{c}$ on diffusion coefficient $D$ and flow velocity $v_0$ in \eqref{eq:49} is highlighted in the transfer function. In \eqref{eq:60a}, the particle concentration $P$ is expressed in terms of transfer function $\bar{\bm{H}}$ which is excited in the transform domain by an input signal, i.e., by external sources $\bar{\bm{F}}_\mathrm{e}$ and initial conditions $\bar{\bm{y}}_\mathrm{init}$. 
Hereby, transfer function $\bar{\bm{H}}$ models the influence of the cylindrical channel on the injected particles, i.e., their propagation based on advection and diffusion. 
The representation in \eqref{eq:60a} is a compact description of the advection-diffusion process in the frequency domain and also allows an analysis of the process in terms of its spectrum. 
Transfer functions of the form of \eqref{eq:60a} are well known in linear operator and control theory, where they are referred to as resolvent operators that are used to study the spectral properties of linear operators \cite{Deutscher:book:2012, Schaefer2020}. 

\subsection{Discrete-time Transfer Function Model}
\vspace*{-1ex}
While the previously derived representations in terms of the CGF \eqref{eq:58} and the transfer function \eqref{eq:60a} provide compact descriptions in the continuous-time domain and the frequency domain, respectively, we also derive a representation in the discrete-time domain for numerical evaluation.
To this end, an impulse-invariant transformation \cite{girod:1997} is applied to state equation \eqref{eq:49} and output equation \eqref{eq:39} to obtain a representation in the discrete-time domain 
\begin{align}
\bar{\bm{y}}^\mathrm{d}[k] &= \As_\mathrm{c}^\mathrm{d}(v_0)\bar{\bm{y}}^\mathrm{d}[k - 1] 
+ \bar{\bm{f}}^\mathrm{d}_\mathrm{e}[k] + \bar{\bm{y}}_\mathrm{init}\delta[k],
&&\As_\mathrm{c}^\mathrm{d}(v_0) = \expE{\As_\mathrm{c}(v_0)T},
\label{eq:60}\\
p^\mathrm{d}[\bm{x},k] &= \bm{c}_1\tran(\bm{x})\bar{\bm{y}}^\mathrm{d}[k],\label{eq:61}
\end{align}
where $t = kT$, $T$ is the  sampling interval, and discrete-time state matrix $\As_\mathrm{c}^\mathrm{d}$ is defined in terms of a matrix exponential. Variables in the discrete-time domain are indicated by superscript $(\cdot)^\mathrm{d}$ and $\delta[k]$ denotes a delta impulse in the discrete-time domain.


\section{Numerical Evaluation}
\label{sec:simul}
\vspace*{-1ex}
In this section, the proposed analytical model is numerically evaluated, i.e., its accuracy is verified by PBS, and the results are compared to existing solutions for the flow dominant and dispersive regimes \cite{jamali:ieee:2019, wicke:globecom:2018, Aris:1956}. Supplementary material including videos and figures is provided in \cite{supplementary}.
\vspace*{-2ex}
\subsection{Simulation Parameters}
\vspace*{-1ex}
The proposed model has been derived and is evaluated in terms of normalized physical quantities with respect to a reference length $\rho$ and a reference time~$\tau$, and therefore, the model can be applied to problems at different scales, i.e., nano, micro, or macro scale. For numerical evaluation, the parameters in Table~\ref{tab:param} have been used, which may model, e.g., micro-fluidic ducts \cite{wicke:globecom:2018, Bruus:2007}, but can also be scaled to model small capillaries \cite{Probstein:Book:2005}. In the following, all parameters, except the diffusion coefficient $D$, are kept constant.
%
For all numerical evaluations, the discrete-time SSD \eqref{eq:60}, \eqref{eq:61} with sampling interval $T = 2\cdot 10^{-4}\,\si{\second}$ was employed, and the number of eigenvalues $Q$ was chosen as
\begin{align}
	&Q = (2N + 1)\cdot M \cdot L, 
	&\bm{q} = \left[ N,\, M,\, L\right],
	\label{eq:65}
\end{align}
where $N$ denotes the maximum orders of Bessel functions $J_n$ used in \eqref{eq:29}, \eqref{eq:30}, i.e., $n = -N, \dots, N$, $M$ is the number of roots $k_{n,m}$ in \eqref{eq:30} for each order $n$, i.e., $m = 0, \dots M-1$, and $L$ denotes the number of wave-numbers $\lambda_\nu$ in \eqref{eq:31}, i.e., $\nu = 0, \dots, L-1$. The selection of the values of $N$, $M$, and $L$ in \eqref{eq:65} directly affects the accuracy of the proposed model and is discussed in detail in Section~\ref{subsec:accuracy}. 

\begin{table}
\caption{Physical parameters for numerical evaluation}
\label{tab:param}
\vspace{-2ex}
\begin{tabular}{p{4cm}p{5cm}p{4.5cm}p{2.5cm}}
\hline\noalign{\smallskip}
Parameter & Value &  Normalized value\\
\noalign{\smallskip}\hline\noalign{\smallskip}
Radius $R_0$& $100\,\si{\micro\meter}$ & $1$	\smallskip\\
Length $Z_0$& $1\,\si{\milli\meter}$ & $10$ \smallskip\\
Flow velocity $v_0$ & $50\,\si{\micro\meter\per\second}$ & 
$50$\smallskip\\
TX/RX distance $d$ & $100\,\si{\micro\meter}$ & 
$1$\smallskip\\
Diffusion coefficient $D$ & $2.5\cdot 10^{-12}\,\si{\square\meter\per\second}\,\dots\, 
5\cdot 10^{-9}\,\si{\square\meter\per\second}
$ & 
$2.5\cdot 10^{-2}\,\dots\,50$\\
\noalign{\smallskip}\hline\noalign{\smallskip}
Reference length $\rho$ & $R_0$ & \smallskip\\
Reference time $\tau$ & $1\cdot 10^{2}\,\si{\second}$ & \\
\noalign{\smallskip}\hline\noalign{\smallskip}
\end{tabular}
\vspace*{-4ex}
\end{table}

\vspace*{-1ex}
\subsection{Initial Conditions}
\vspace*{-1ex}
For the analysis and numerical evaluation of the proposed model, we consider two different initial distributions of the particles in the cylinder, i.e., a uniform distribution and a point distribution. For the initial distributions, the following raised cosine function is defined 
\begin{align}
	f_\mathrm{i}(\chi, \chi_0, \chi_\mathrm{e}) = 
	\begin{cases}
		\frac{1}{2}\left(1 +\cos(\frac{2\pi}{\chi_0}(\chi - \chi_\mathrm{e})) \right) & \chi_\mathrm{e}-\frac{\chi_0}{2} \le \chi \le \chi_\mathrm{e}+\frac{\chi_0}{2}\\
		0 & \mathrm{else}
	\end{cases},\label{eq:65}
\end{align}
with a spatial width $\chi_0$ and center position $\chi_\mathrm{e}$.\footnote{We note that instead of the raised cosine function \eqref{eq:65}, any other smooth function can be used to model the initial distribution of particles.}

\subsubsection{Uniform Distribution} 
In the considered uniform distribution, the particles are uniformly distributed in the $r$-$\varphi$-plane of the cylinder. In $z$-direction, the initial distribution of particles is centered at $z = z_\mathrm{e}$ and spread over $z_0$ as defined by \eqref{eq:65}.
The 3D uniform distribution is defined by specifying initial conditions $p_\mathrm{init}$ in \eqref{eq:11} as follows
\begin{align}
	p_\mathrm{init}(\bm{x}) = p_\mathrm{uniform}(\bm{x}) 
	\coloneqq f_{\mathrm{i}}(z, z_0, z_\mathrm{e}), \label{eq:66}
\end{align}
with normalized width $z_0 = 0.3$ and normalized center position $z_\mathrm{e} = 1$. 
The considered uniform distribution is shown in the plots on the left hand side of Fig.~\ref{fig:uniform2D}.
\subsubsection{Point Distribution} 
Furthermore, a point distribution centered at $\bm{x}_{\mathrm{TX}} = \left[r_\mathrm{e}, \varphi_\mathrm{e}, z_\mathrm{e} \right]$ is considered. The particles are distributed as defined in \eqref{eq:65} in all spatial directions. The 3D point distribution is defined by specifying initial conditions $p_\mathrm{init}$ in \eqref{eq:11} as follows
\begin{align}
	p_\mathrm{init}(\bm{x}) = 
	p_\mathrm{point}(\bm{x}) \coloneqq 
	f_{\mathrm{i}}(r, r_0, r_\mathrm{e})\,  f_{\mathrm{i}}(\varphi, \varphi_0, \varphi_\mathrm{e})\, f_{\mathrm{i}}(z, z_0, z_\mathrm{e}), \label{eq:67}
\end{align}
with normalized widths $r_0 = 0.4,\,\varphi_0 = \frac{\pi}{4},\, z_0 = 0.4$. The center positions in $\varphi$- and $z$-direction are $\varphi_\mathrm{e} = \frac{\pi}{2}$ and $z_\mathrm{e} = 1$, respectively, while the radial center position is varied as $r_\mathrm{e} = 0.25,\, 0.5, \, 0.75$. 
The considered point distribution is shown in the plots on the left hand side of Fig.~\ref{fig:point2D}.

Mostly, MC channel models are derived and analyzed by assuming a point release of particles that is defined in terms of $\delta$-impulses, see e.g. \cite[Eq.~(11)]{Zoofaghari:ieee:2019}, \cite[Eq.~(5d)]{schaefer:icc:2019} for cylindrical environments. 
However, such impulsive releases are unrealistic, as in practical systems particles cannot be released from an infinitesimal point, but the assumption simplifies the derivation of the channel model and the channel impulse response. To go one step towards more realistic channel models, we consider a point release of particles over a non-zero volume \eqref{eq:67}.
Another benefit of the adopted release profile is the suppression of Gibbs phenomenon which otherwise may occur for all modeling techniques based on modal expansions or CGFs \cite{girod:1997}.

\subsection{Validation Parameters}
\vspace*{-1ex}
For validation, we employ PBS of the considered advection-diffusion process. For PBS, the concentration is estimated by counting the number of observed particles in a cuboid volume $V_\mathrm{cube} = 0.04\times 0.04 \times 0.04$ centered at receiver position $\bm{x}_\mathrm{RX}$. 
For the uniform release scenario in Section~\ref{subsec:sim:uni}, we released $N_\mathrm{TX} = 1\cdot 10^{3}$ particles and their positions are updated in discrete time steps $\Delta t = 10^{-2}\,\si{\second}$. The PBS results are averaged over $8000$ realizations of the process. 
For the point release in Section~\ref{subsec:sim:point}, $N_\mathrm{TX} = 1\cdot 10^{3}$ particles are released and their positions are updated with $\Delta t = 5\cdot 10^{-3}\,\si{\second}$ and the results are averaged over $50\cdot 10^{3}$ realizations.

 The proposed model is evaluated at point $\bm{x}_\mathrm{RX}$, which is the center of the cuboid used for PBS. 
 Using the uniform concentration assumption in the cuboid \cite{noel:ieee:2013}, output equation \eqref{eq:61} becomes
\begin{align}
	&p^\mathrm{d}_{\mathrm{cube}}[k] = V_\mathrm{cube}\cdot \bm{c}_1\tran(\bm{x}_\mathrm{RX})\,\bar{\bm{y}}^\mathrm{d}[k].
	 \label{eq:70}
\end{align} 
Furthermore, the proposed model is classified with respect to existing analytical models. Therefore, the dispersion factor $\alpha$, is introduced \cite[Eq.~(20)]{jamali:ieee:2019}  
\begin{align}
	\alpha = \frac{D\,d}{v_{\mathrm{eff}} R_0^2}
	= \frac{2 D\,d}{v_{0} R_0^2}
	\label{eq:68}
\end{align}
to distinguish between different regimes.
As summarized in \cite[Sec.~D-2]{jamali:ieee:2019} the transport of particles in the considered scenarios can be categorized into three regimes, i.e., the flow dominant regime ($\alpha \ll 1$), the dispersive regime ($\alpha \gg 1$), and the mixed regime. Furthermore, there are well-known solutions and approximations for the flow dominant regime, see \cite[Eq.~(16)]{wicke:globecom:2018}, and the dispersive regime, see \cite{jamali:ieee:2019}, \cite[Eq.~(11)]{wicke:globecom:2018}, \cite{Aris:1956}.   

In the following, dispersion factor $\alpha$ is used to classify the considered scenarios into different regimes where the different regimes are realized by changing the diffusion coefficient $D$, see Table~\ref{tab:alpha}. 
%
We note that other parameters may also be varied to evaluate the model in different regimes, see \eqref{eq:68}. Varying the diffusion coefficient $D$ allows, e.g., to analyze the behavior of particles of different sizes in a given channel.


\begin{table}[t]
\caption{\small Considered diffusion coefficients and resulting dispersion factors}
\label{tab:alpha}
\vspace{-1ex}
\centering
\begin{tabular}{p{2cm}|p{1.5cm}p{1.5cm}p{1.5cm}p{1.5cm}p{1.5cm}p{1.5cm}p{1.5cm}}
\hline\noalign{\smallskip}
$D$ in $\si{\square\meter\per\second}$ & $2.5\cdot 10^{-12}$ & $2.5\cdot 10^{-11}$ & $2.5\cdot 10^{-10}$ & $7.5\cdot 10^{-10}$ & $1.25\cdot 10^{-9}$ & $2.5\cdot 10{-9}$ & $5\cdot 10^{-9}$\\
\hline\noalign{\smallskip}
$\alpha$ & $1\cdot 10^{-3}$ & $1\cdot 10^{-2}$ & $0.1$ & $0.3$ & $0.5$ & $1$ & $2$\\
\hline\noalign{\smallskip}
\end{tabular}
\vspace*{-4ex}
\end{table}

\subsection{Uniform Release}
\label{subsec:sim:uni}
\vspace*{-1ex}
In this section, the proposed model is evaluated for a uniform release of particles modeled by initial condition \eqref{eq:66}. For numerical evaluation of the proposed model, a total number of $Q = 6000$ eigenvalues is used, i.e., $\bm{q} = [0,\, 30,\, 200]$. Here, $N = 0$ is due to the initial condition in \eqref{eq:66}, i.e., as the initial distribution is radially symmetrical, only Bessel functions of order $n = 0$ contribute to the solution. The receiver is placed at $\bm{x}_\mathrm{RX} = [0, \nicefrac{\pi}{2}, 2]$, where the concentration is calculated based on \eqref{eq:70}.
%

\begin{figure*}[t]
    \centering
    \begin{minipage}{0.3\textwidth}
    	\centering
        \includegraphics[width=\linewidth]{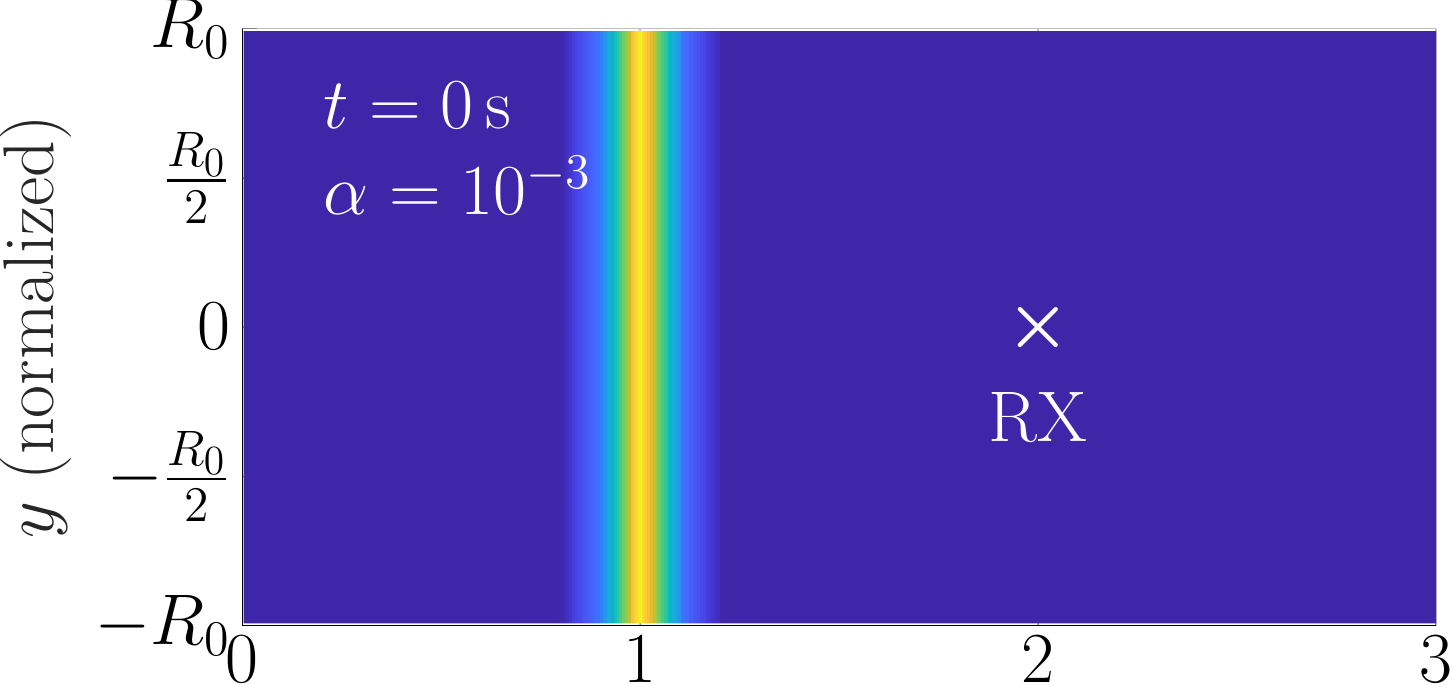}
    \end{minipage}\hfill
    \begin{minipage}{0.3\textwidth}
    	\centering
    	\vspace*{0.5ex}
		\includegraphics[width=0.85\linewidth]{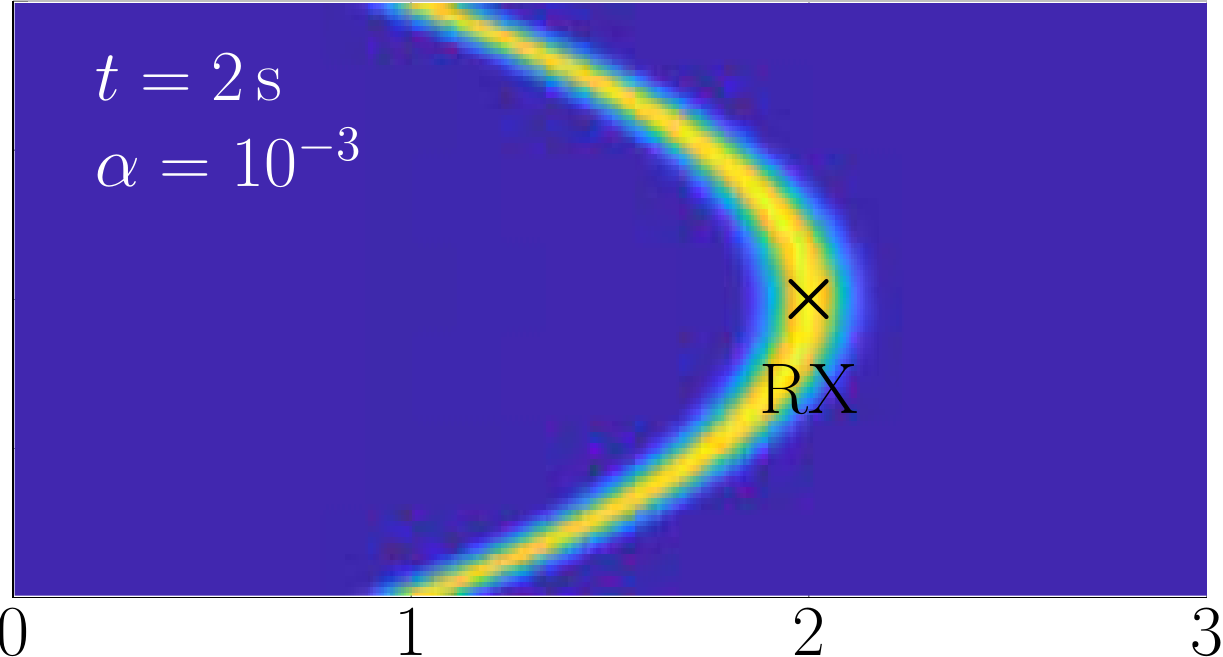}
    \end{minipage}\hfill
    \begin{minipage}{0.3\textwidth}
    	\centering
    	\vspace*{0.5ex}
		\includegraphics[width=0.85\linewidth]{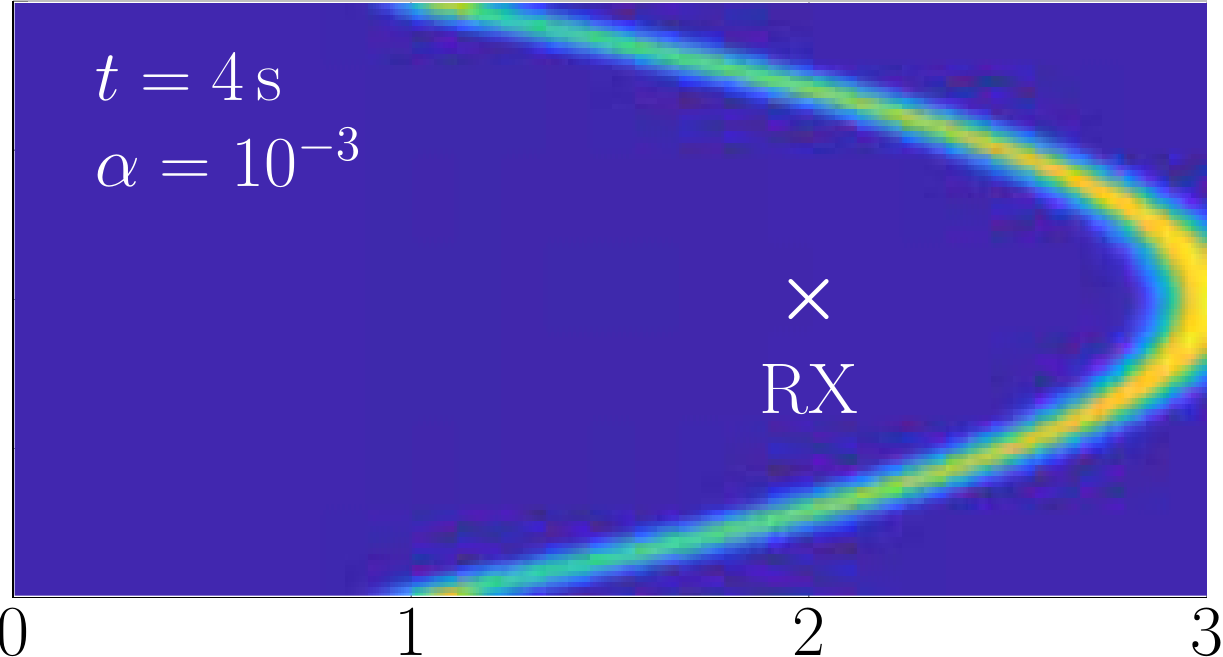}
    \end{minipage}\\
    \begin{minipage}{0.3\textwidth}
    	\centering
        \includegraphics[width=\linewidth]{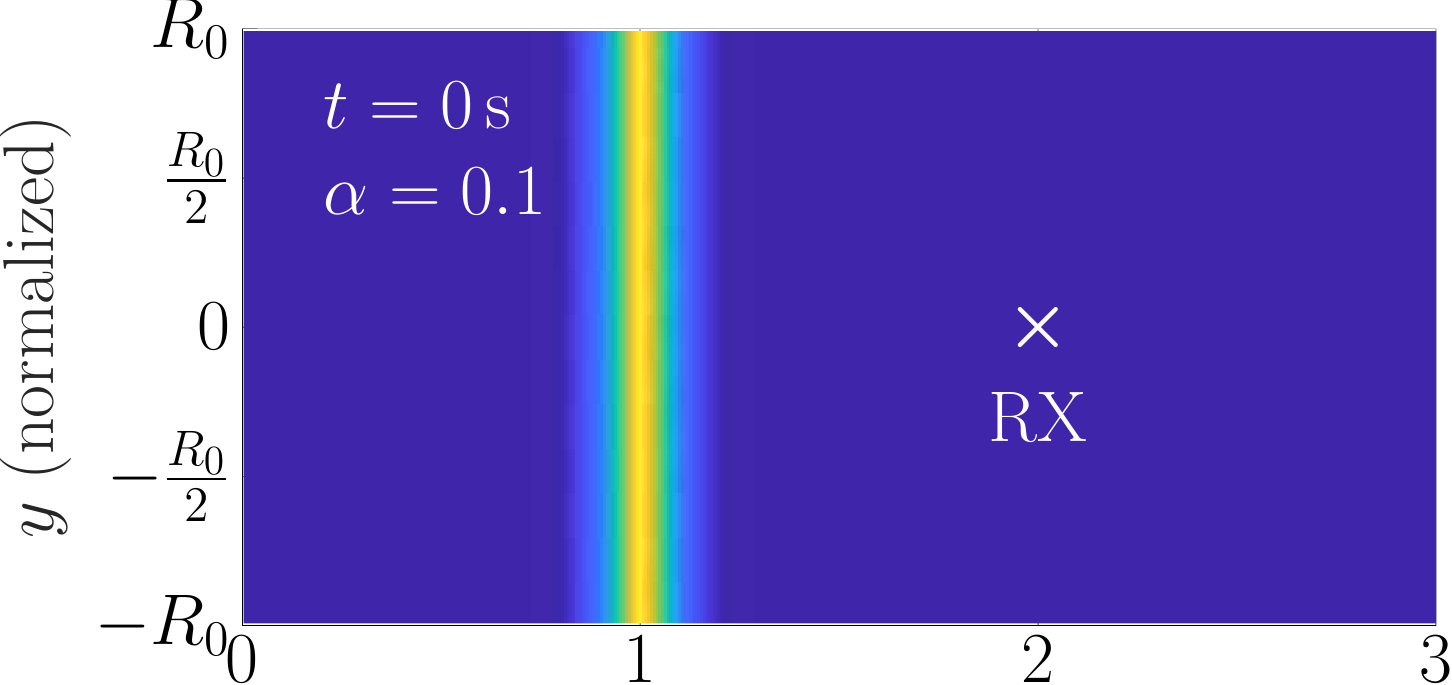}
    \end{minipage}\hfill
    \begin{minipage}{0.3\textwidth}
    	\centering
    	\vspace*{0.5ex}
		\includegraphics[width=0.85\linewidth]{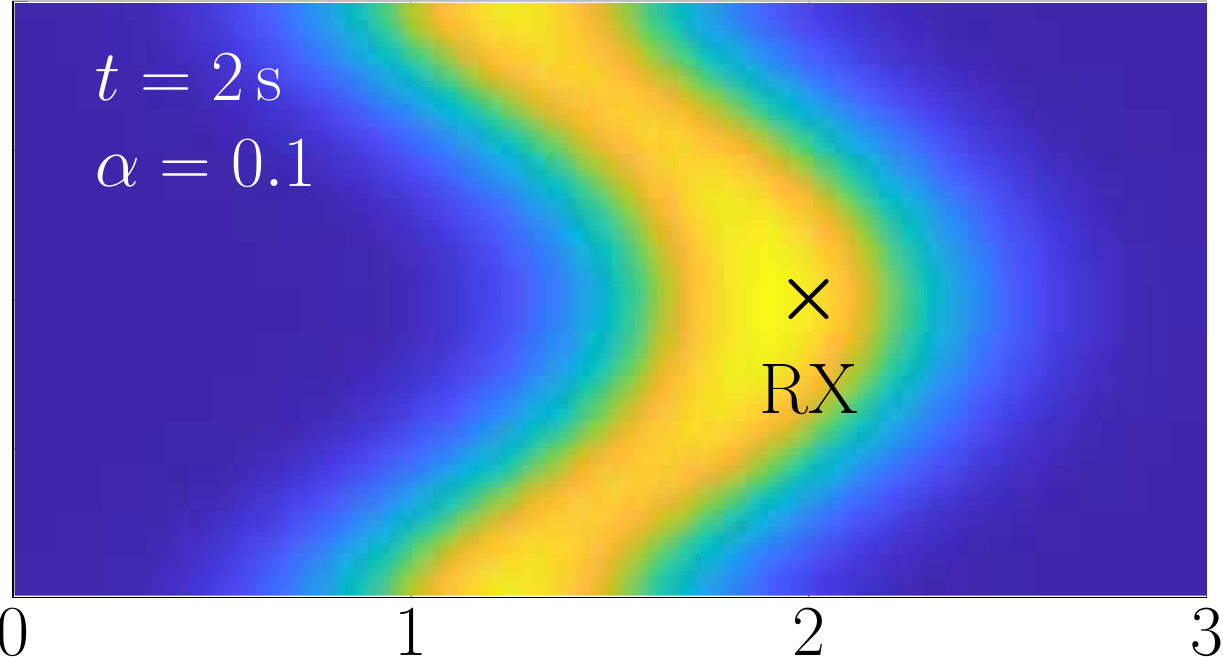}
    \end{minipage}\hfill
    \begin{minipage}{0.3\textwidth}
    	\centering
    	\vspace*{0.5ex}
		\includegraphics[width=0.85\linewidth]{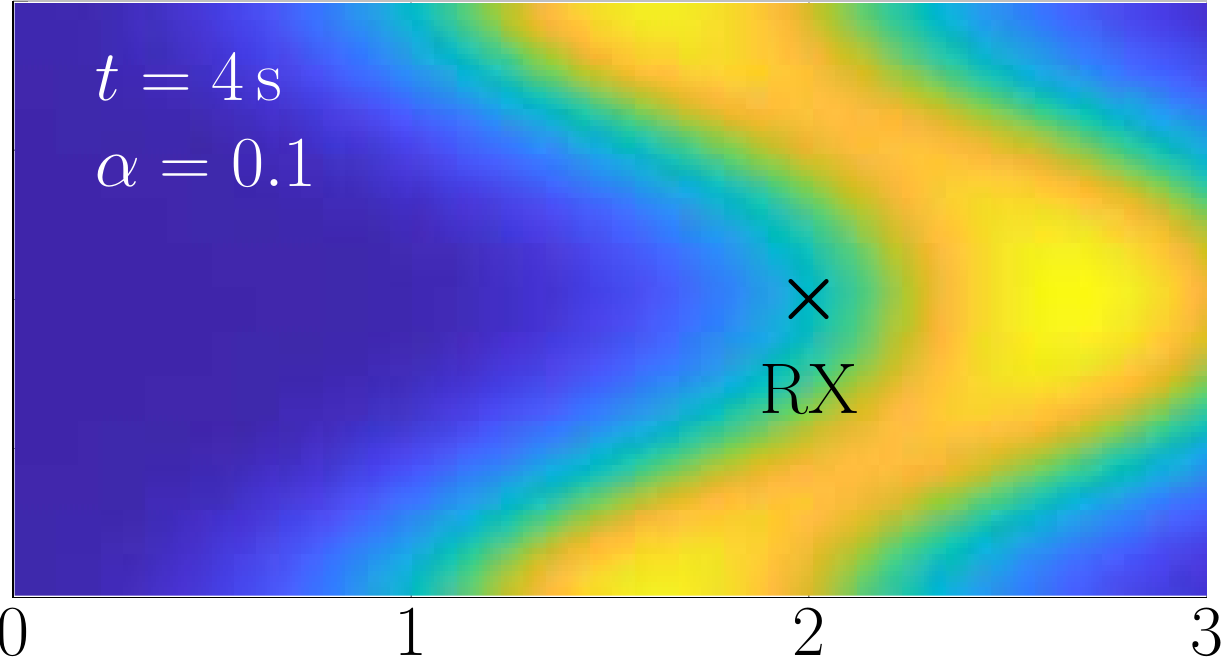}
    \end{minipage}
    \\
    \begin{minipage}{0.3\textwidth}
    	\centering
        \includegraphics[width=\linewidth]{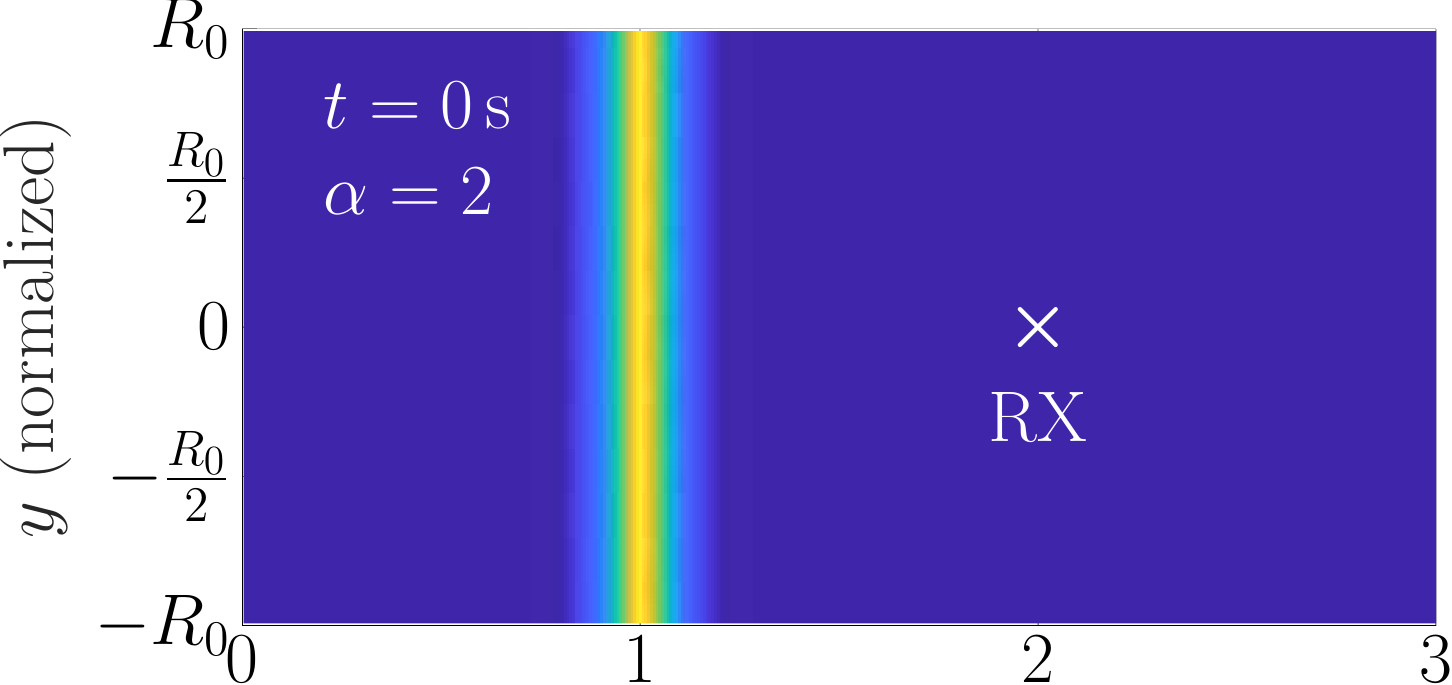}
    \end{minipage}\hfill
    \begin{minipage}{0.3\textwidth}
    	\centering
    	\hspace*{0.01ex}
    	\vspace*{-0.5ex}
		\includegraphics[width=0.85\linewidth]{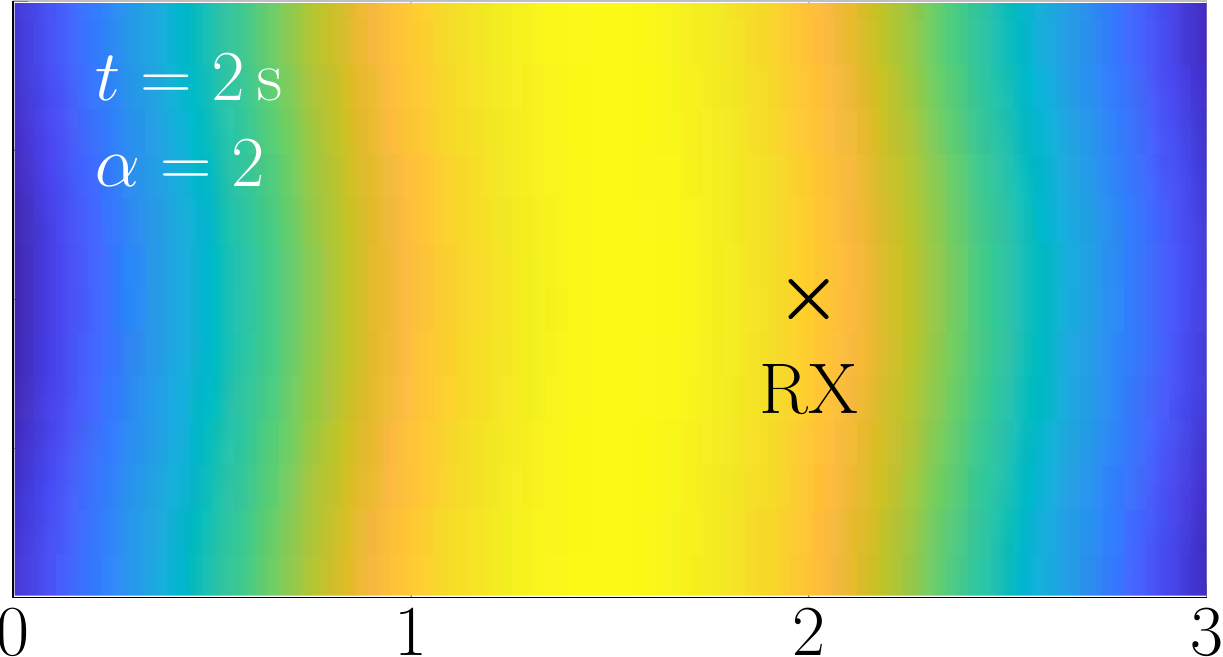}
    \end{minipage}\hfill
    \begin{minipage}{0.3\textwidth}
    	\centering
       	\hspace*{0.02ex}
       	\vspace*{-0.5ex}
		\includegraphics[width=0.85\linewidth]{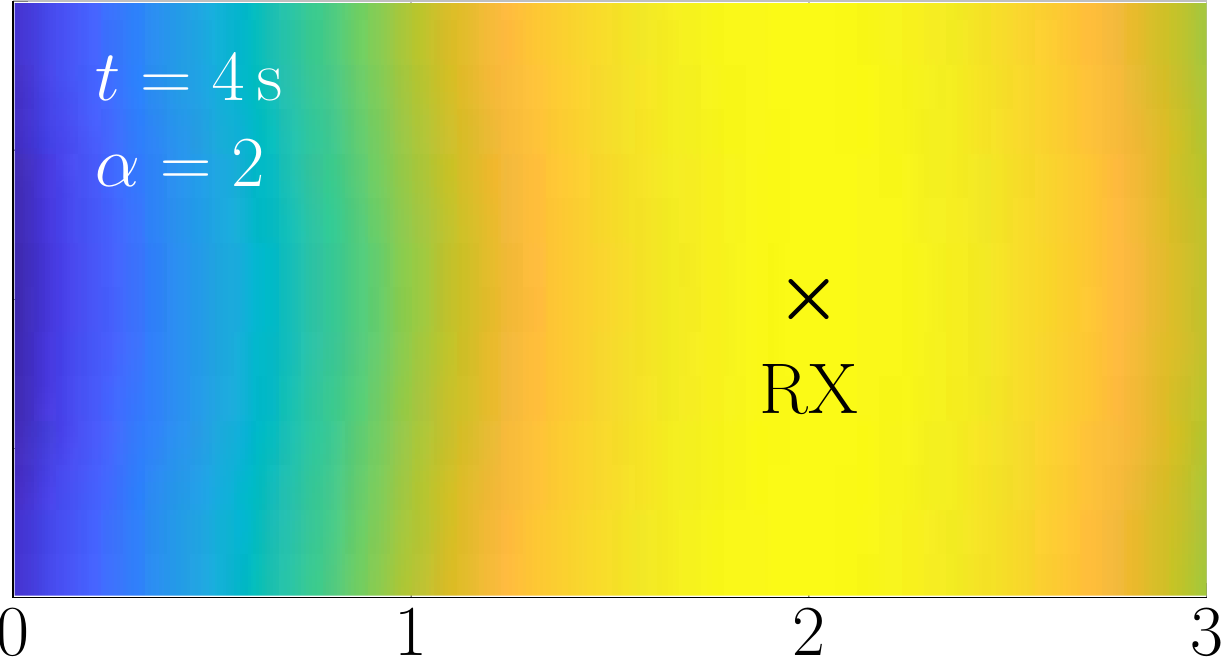}
    \end{minipage}
    \vspace{-2ex}
    \caption{\small 2D concentration $p(\bm{x},t)$ in the $y$-$z$-plane ($y = r\,\sin(\frac{\pi}{2})$) of the cylinder at times $t = 0\,\si{\second},  2\,\si{\second},  4\,\si{\second}$ for a uniform release.
        Different values $\alpha = 1\cdot 10^{-3},\, 0.1,\, 2$ are employed (top to bottom). 
}
    \label{fig:uniform2D}
    \vspace*{-3ex}
\end{figure*}

\begin{figure*}[t]
    \centering
    \begin{subfigure}[b]{0.3\textwidth}
            \centering
            	\includegraphics[width=0.9\linewidth]{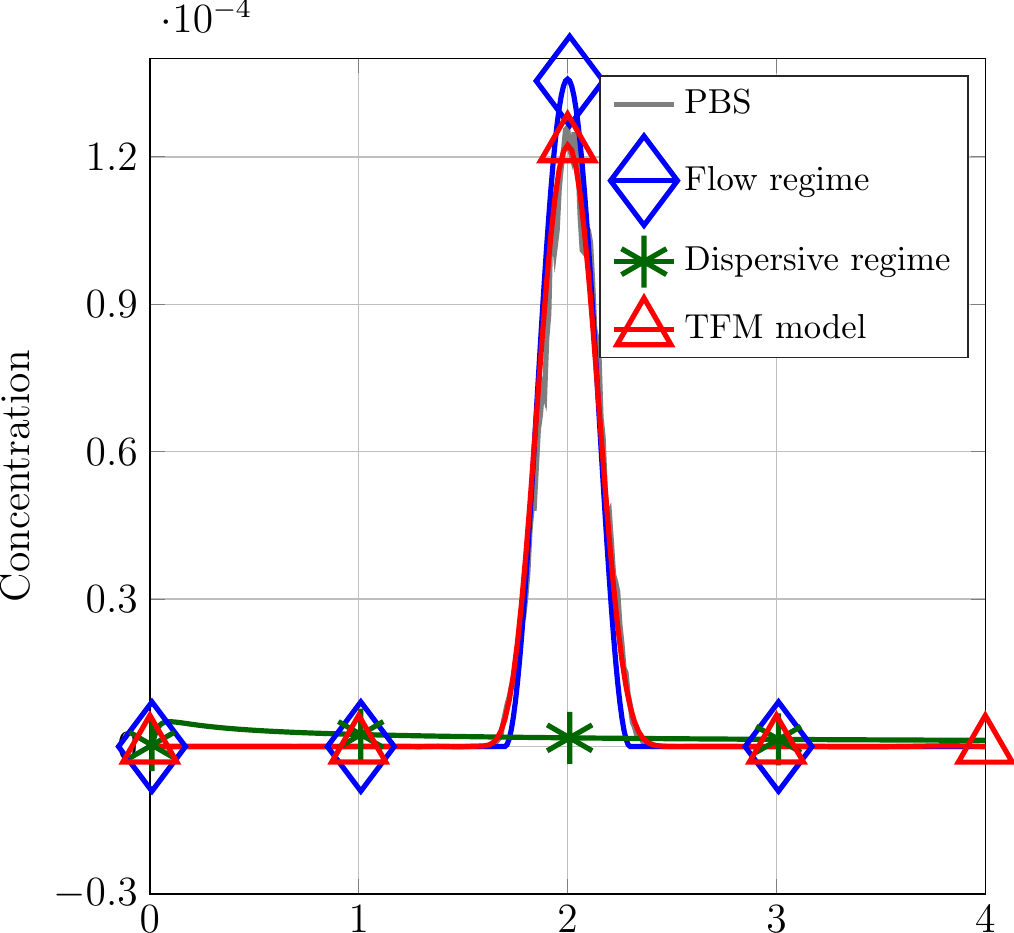}
            	\vspace{-1ex}
            \caption{$\alpha = 1\cdot 10^{-3}$}
    \label{fig:uniform1D:a}
    \end{subfigure}
    \begin{subfigure}[b]{0.3\textwidth}
            \centering
     \includegraphics[width=0.83\linewidth]{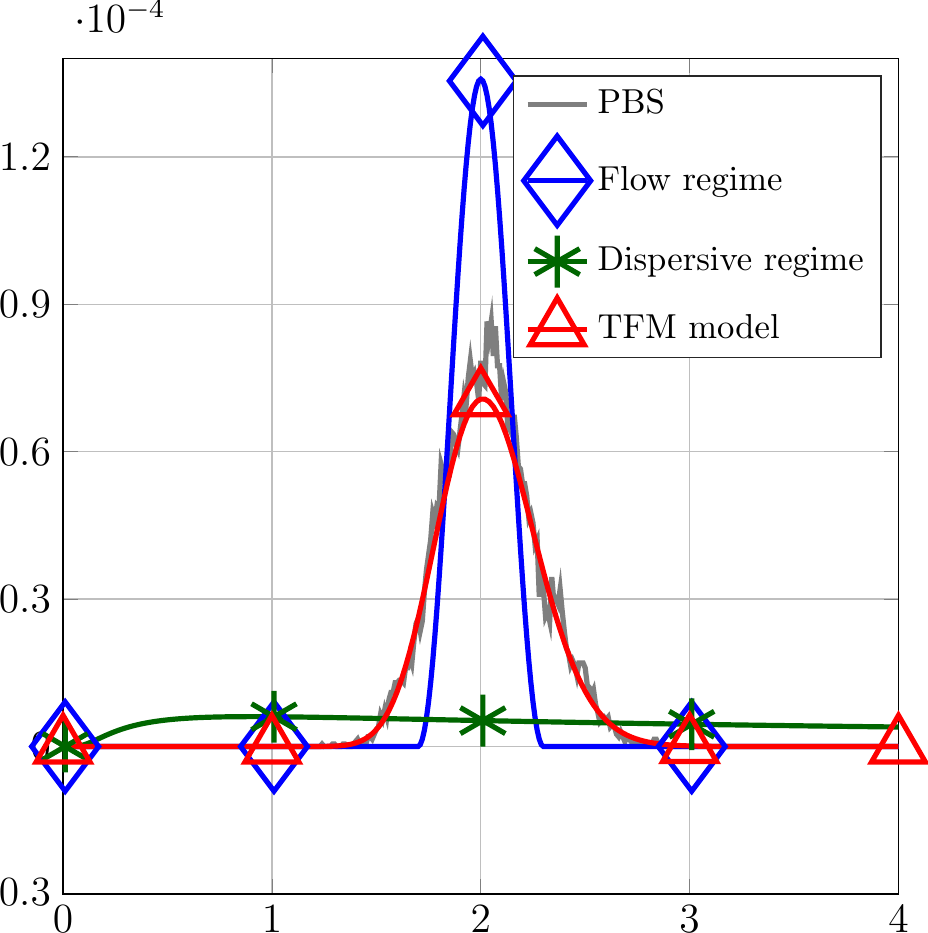}
            	\vspace{-1ex}
            \caption{$\alpha = 1\cdot 10^{-2}$}
    \label{fig:uniform1D:b}
    \end{subfigure}
    \begin{subfigure}[b]{0.3\textwidth}
            \centering
      \includegraphics[width=0.83\linewidth]{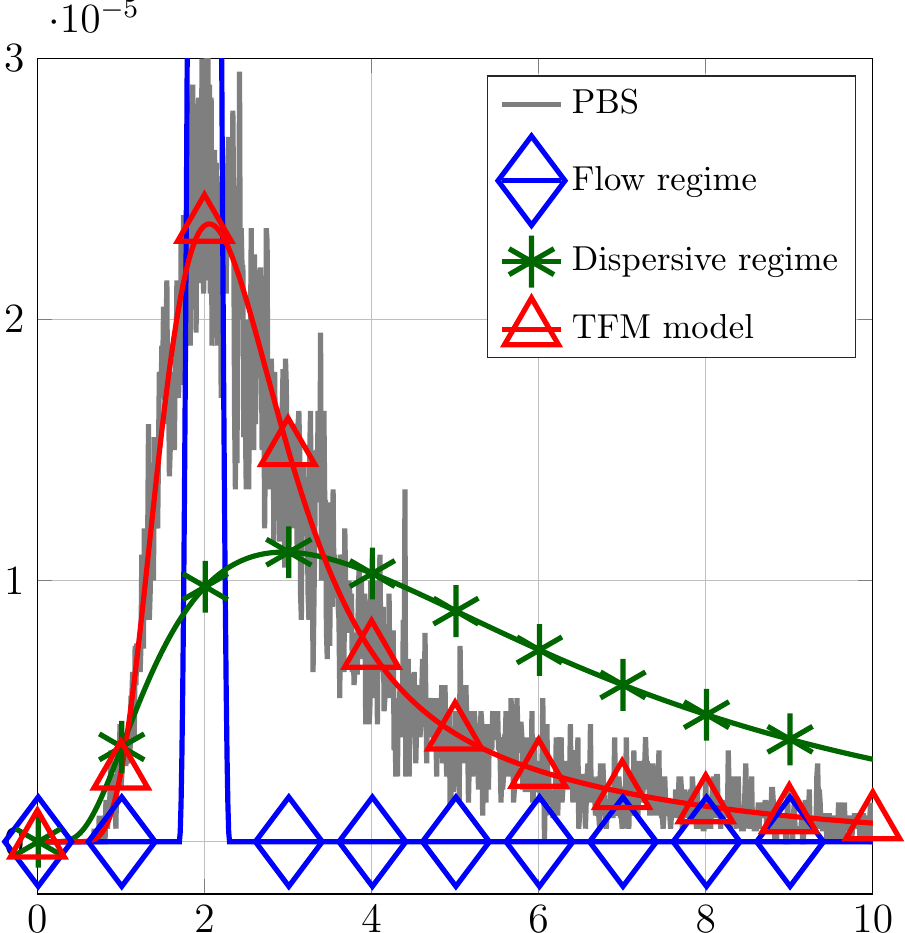}
			\vspace{-1ex}
            \caption{$\alpha = 0.1$}
    \label{fig:uniform1D:c}
    \end{subfigure}\\
    \begin{subfigure}[b]{0.3\textwidth}
            \centering
            \includegraphics[width=0.9\linewidth]{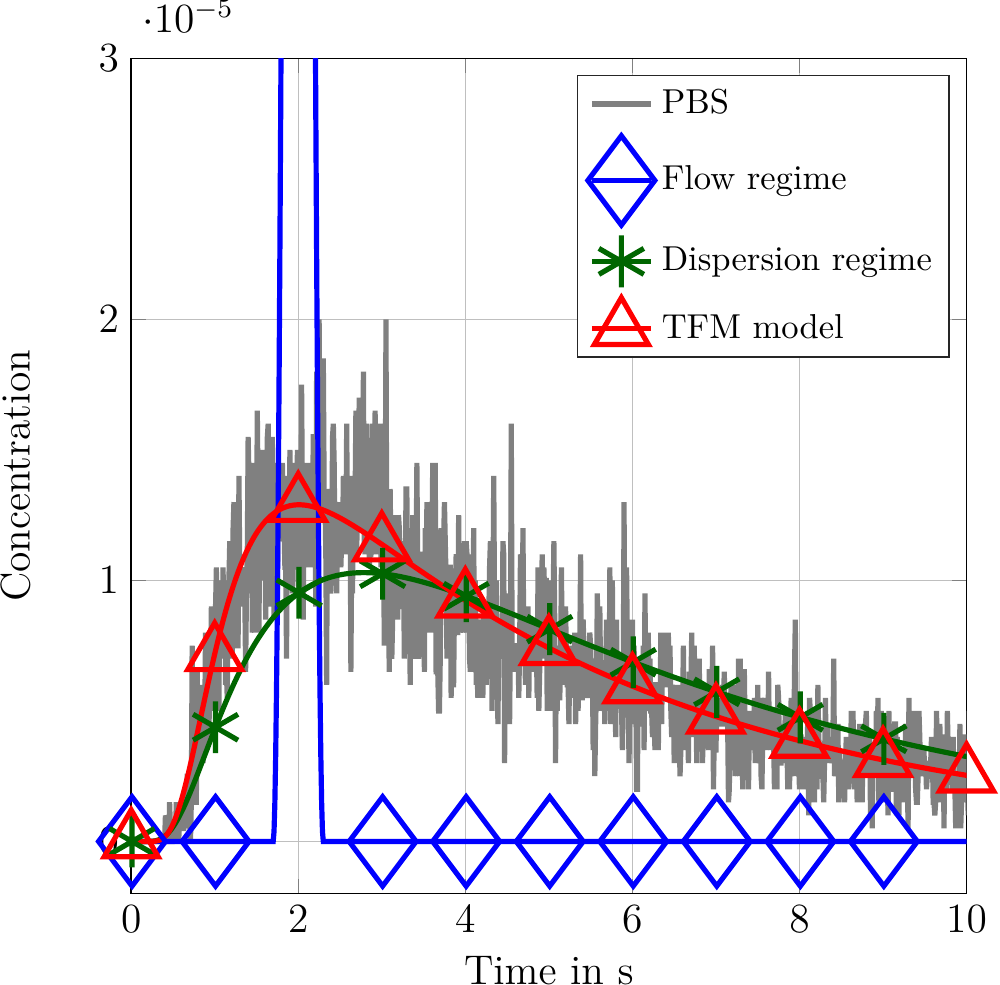}
            	\vspace{-1ex}
            \caption{$\alpha = 0.3$}
    \label{fig:uniform1D:d}
    \end{subfigure}
    \begin{subfigure}[b]{0.3\textwidth}
            \centering
            	\includegraphics[width=0.83\linewidth]{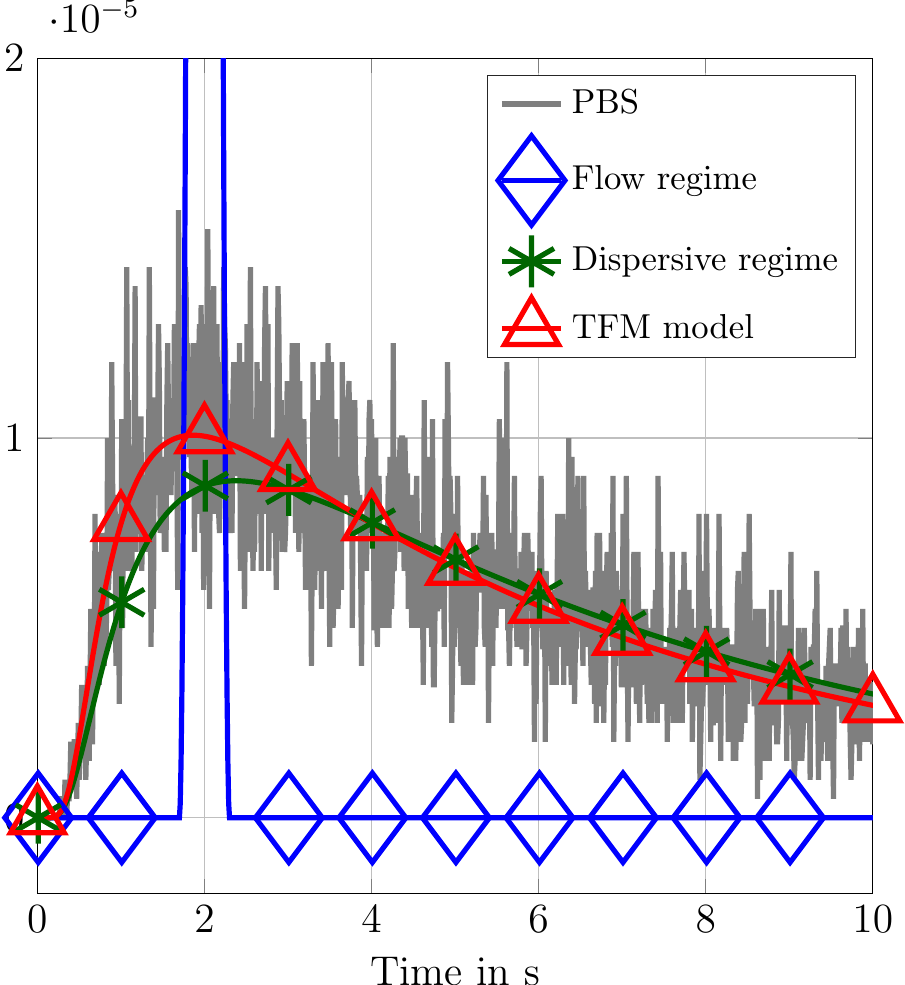}
            	\vspace{-1ex}	
            \caption{$\alpha = 0.5$}
    \label{fig:uniform1D:e}
    \end{subfigure}
    \begin{subfigure}[b]{0.3\textwidth}
            \centering
            \includegraphics[width=0.83\linewidth]{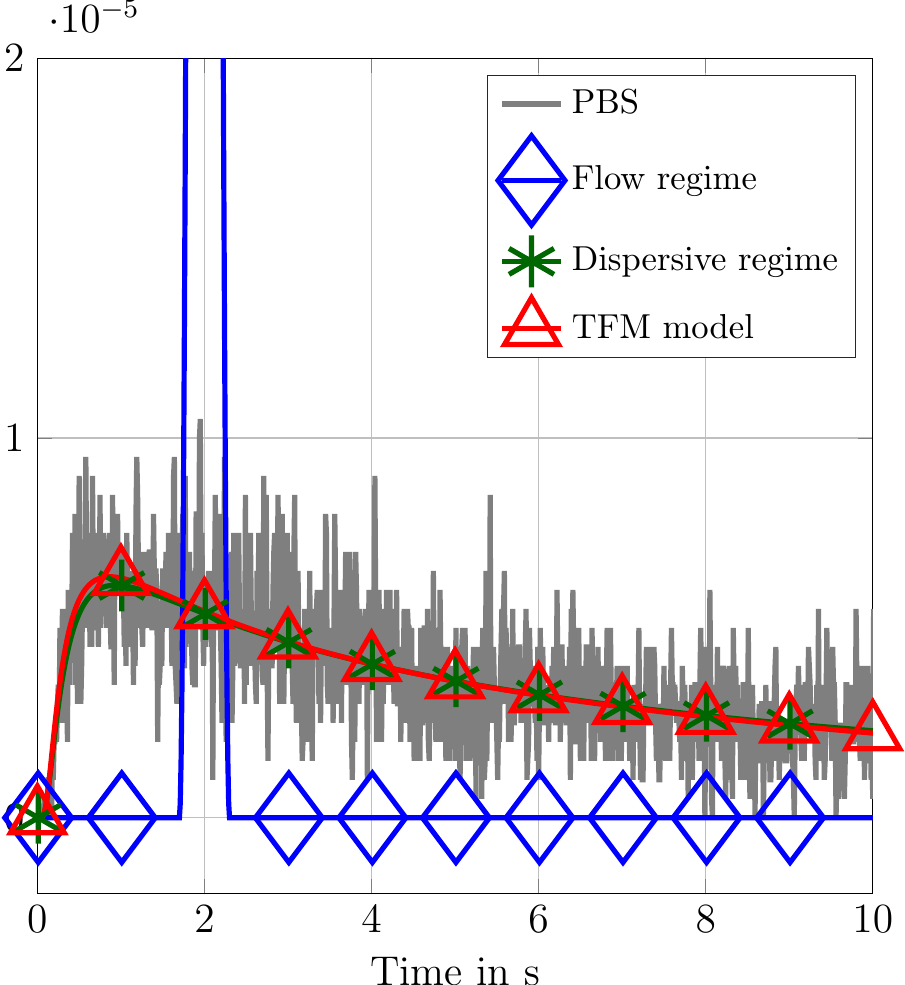}
            	\vspace{-1ex}
            \caption{$\alpha = 2$}
    \label{fig:uniform1D:f}
    \end{subfigure}
    \vspace{-2ex}
    \caption{\small Concentration $p^{\mathrm{d}}_\mathrm{cube}$ at $\bm{x}_\mathrm{RX}$ over time for a uniform release of particles \eqref{eq:66}. Different values of $\alpha$ are considered.}
    \label{fig:uniform1D}
    \vspace*{-3.5ex}
\end{figure*}

 In Fig.~\ref{fig:uniform2D}, the concentration of the particles after a uniform release is presented for the flow dominant ($\alpha = 1\cdot 10^{-3}$, top row), mixed ($\alpha = 0.1$, center row), and dispersive ($\alpha = 2$, bottom row) regime at times $t = 0\,\si{\second},  2\,\si{\second}$, and $4\,\si{\second}$ of the process. 
 The figure illustrates the differences between the defined regimes. 
 In the flow dominant regime (see top row of Fig.~\ref{fig:uniform2D}), the influence of flow is dominant and diffusion has little impact. 
 The characteristic parabolic profile of the laminar flow, see \eqref{eq:5}, becomes obvious over time, with maximum velocity $v(0) = v_0$ in the center of the cylinder and zero velocity $v(R_0) = 0$ at the boundaries. 
 The spatial spreading of the initial distribution is preserved at the considered RX position (see also Fig.~\ref{fig:uniform1D:a}). 
 In the mixed regime (see center row of Fig.~\ref{fig:uniform2D}), the flow profile is blurred by diffusion. 
 In fact, both flow and diffusion influence the propagating particles.
 With increasing distance from the TX position, the initial uniform distribution is spread over space. 
 This effect becomes even stronger in the dispersive regime (see bottom row of Fig.~\ref{fig:uniform2D}), where the impact of diffusion dominates the impact of flow on the propagating particles.
 The initial distribution of the particles is noticeably spread over space already after $t = 2\,\si{\second}$ (see second plot in bottom row of Fig.~\ref{fig:uniform2D}). 
  
Fig.~\ref{fig:uniform1D} shows the particle concentration at RX position $\bm{x}_\mathrm{RX}$ for different values of $\alpha$. The figure shows results for the numerical evaluation of the proposed model (red color) and PBS as a ground truth (gray color). Furthermore, the existing solutions for the flow dominant (blue color) and dispersive (green color) regimes are shown. The most important observation from Fig.~\ref{fig:uniform1D} is that the proposed model perfectly matches the PBS results for all considered regimes, which underlines the ability of the model to provide a  solution valid for all regimes. 
In both limiting cases, the existing solutions for the flow dominant regime (Fig.~\ref{fig:uniform1D:a}) and the dispersive regime (Fig.~\ref{fig:uniform1D:f}) also provide a good estimate for the received concentration.  
The plots in Fig.~\ref{fig:uniform1D} highlight the influence of the different regimes on the propagation of the particles, and reinforce the observations made in Fig.~\ref{fig:uniform2D}. 
The peakiness of the uniform particle release profile is evident in the flow dominant regime (Fig.~\ref{fig:uniform1D:a}), and the temporal width $t_\mathrm{peak}$ of the released concentration can be related to the spatial width of the uniform release, i.e., $t_\mathrm{peak}\approx \frac{z_0}{v_0} = 0.6\,\si{\second}$. 
For $\alpha = 1\cdot 10^{-2}$ and $0.1$ (Figs~\ref{fig:uniform1D:b}, \subref{fig:uniform1D:c}) the peak is spread by diffusion, but still recognizable. In both figures, the mismatch between the known solutions for the limiting regimes and the results from PBS are obvious.
The effect of diffusion starts to become dominant for $\alpha = 0.3$ and $0.5$ in Figs.~\ref{fig:uniform1D:d}, \subref{fig:uniform1D:e}. 
In this case, the tail of the received concentration increases, which is directly related to the spatial spreading of the uniform release, see center row of Fig.~\ref{fig:uniform2D}. Furthermore, both figures show that for larger values of $\alpha$ the known solution for the dispersive regime (green color) starts to provide a better estimate for the received concentration.
For $\alpha = 2$ (dispersive regime) in Fig.~\ref{fig:uniform1D:f}, diffusion clearly dominates. The peak is completely spread and, compared to the other scenarios in Fig.~\ref{fig:uniform1D}, even the temporal location of the maximum received concentration occurs earlier due to the high diffusion. In this scenario, the known solution for the dispersive regime provides a good estimate for the concentration.


\subsection{Point Release}
\label{subsec:sim:point}
\vspace*{-1ex}
In this section, a point release of the form in \eqref{eq:67} is considered. For numerical evaluation $Q = 12.6\cdot 10^{4}$ is used, i.e., $\bm{q} = [20, 30, 200]$. 
Because of the point release,
a large value of $N$ is necessary to correctly represent the propagation of the particles in the 3D volume, see Section~\ref{subsec:accuracy}. Due to the symmetry of Bessel functions, i.e., $J_{-n}(x) = (-1)^{n}\,J_n(x)$, the negative values of $n = -N, \dots, -1$ are not evaluated separately, which reduces \eqref{eq:65} to $Q = (N+1)\cdot M\cdot L$. 
Point releases at three different release positions with $r_\mathrm{e} = 0.25,\, 0.5$, and $0.75$ are considered, while the receiver position is fixed at $\bm{x}_\mathrm{RX} = [0.5, \,\nicefrac{\pi}{2},\, 2]$. 
%
%

\begin{figure*}[t]
    \centering
    \begin{minipage}{0.3\textwidth}
    	\centering
        \includegraphics[width=\linewidth]{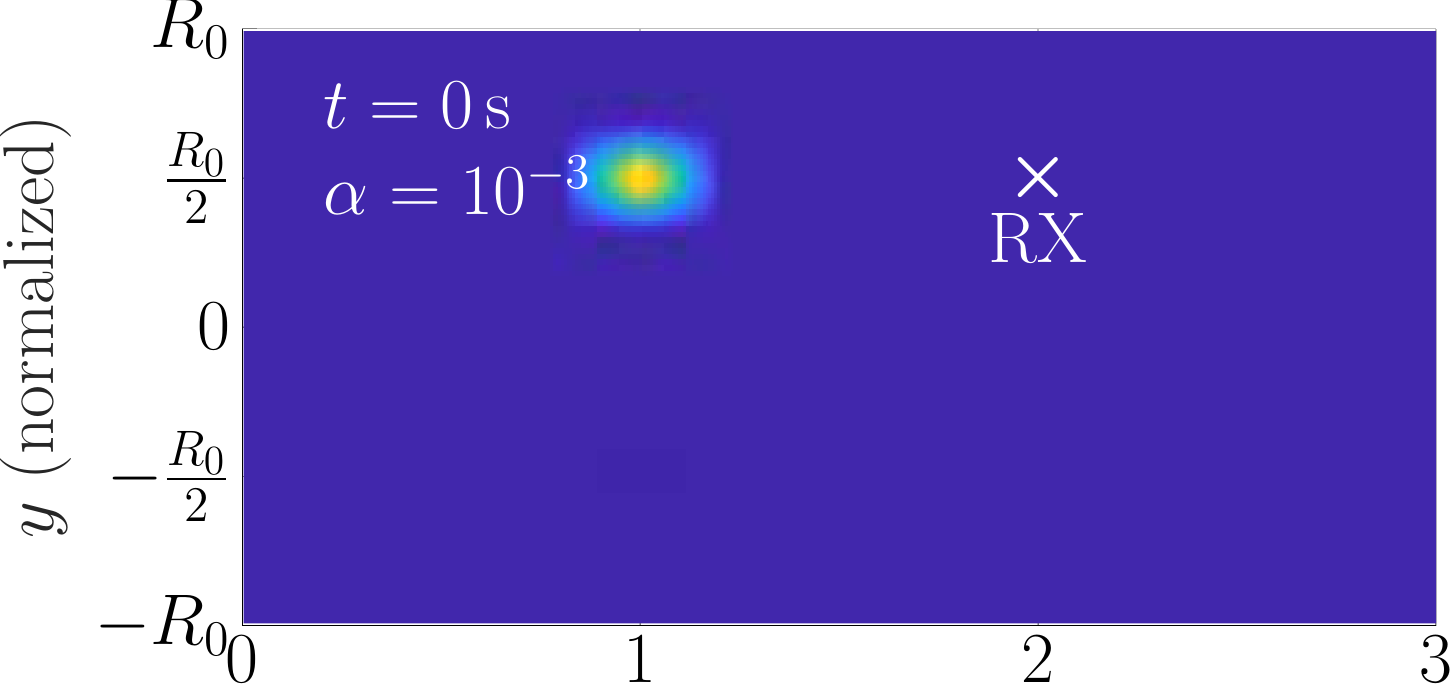}
    \end{minipage}\hfill
    \begin{minipage}{0.3\textwidth}
    	\centering
    	\vspace*{0.5ex}
	\includegraphics[width=0.85\linewidth]{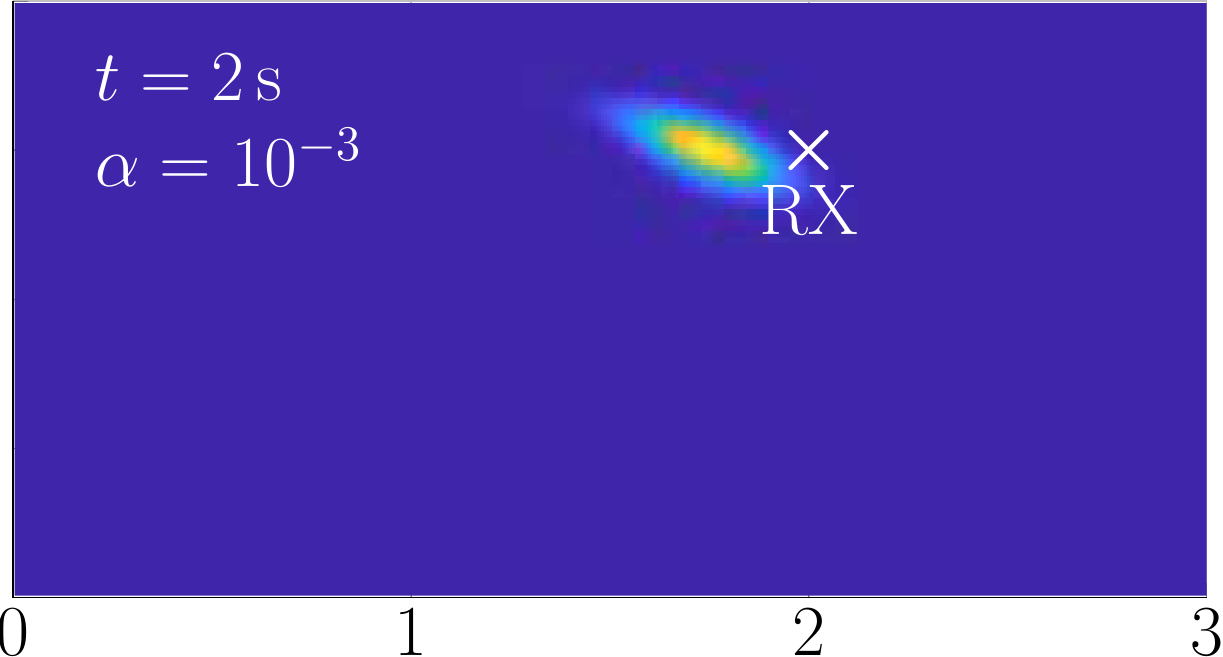}
    \end{minipage}\hfill
    \begin{minipage}{0.3\textwidth}
    	\centering
    	\vspace*{0.5ex}
	\includegraphics[width=0.85\linewidth]{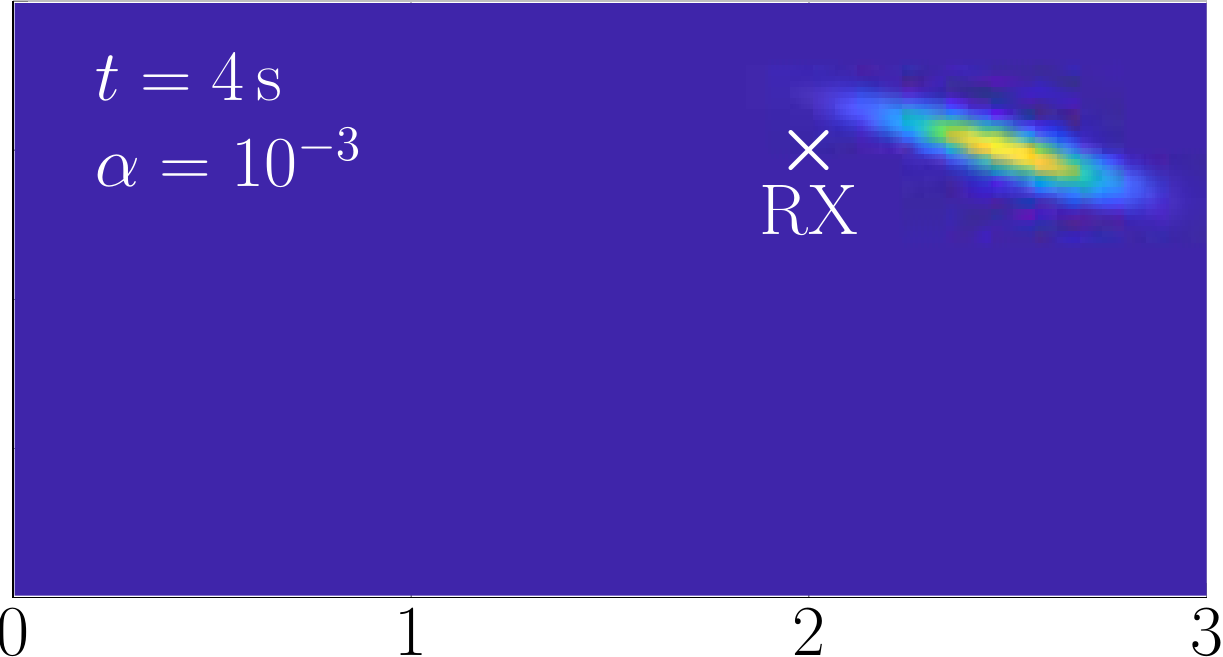}
    \end{minipage}\\
    \begin{minipage}{0.3\textwidth}
    	\centering
        \includegraphics[width=\linewidth]{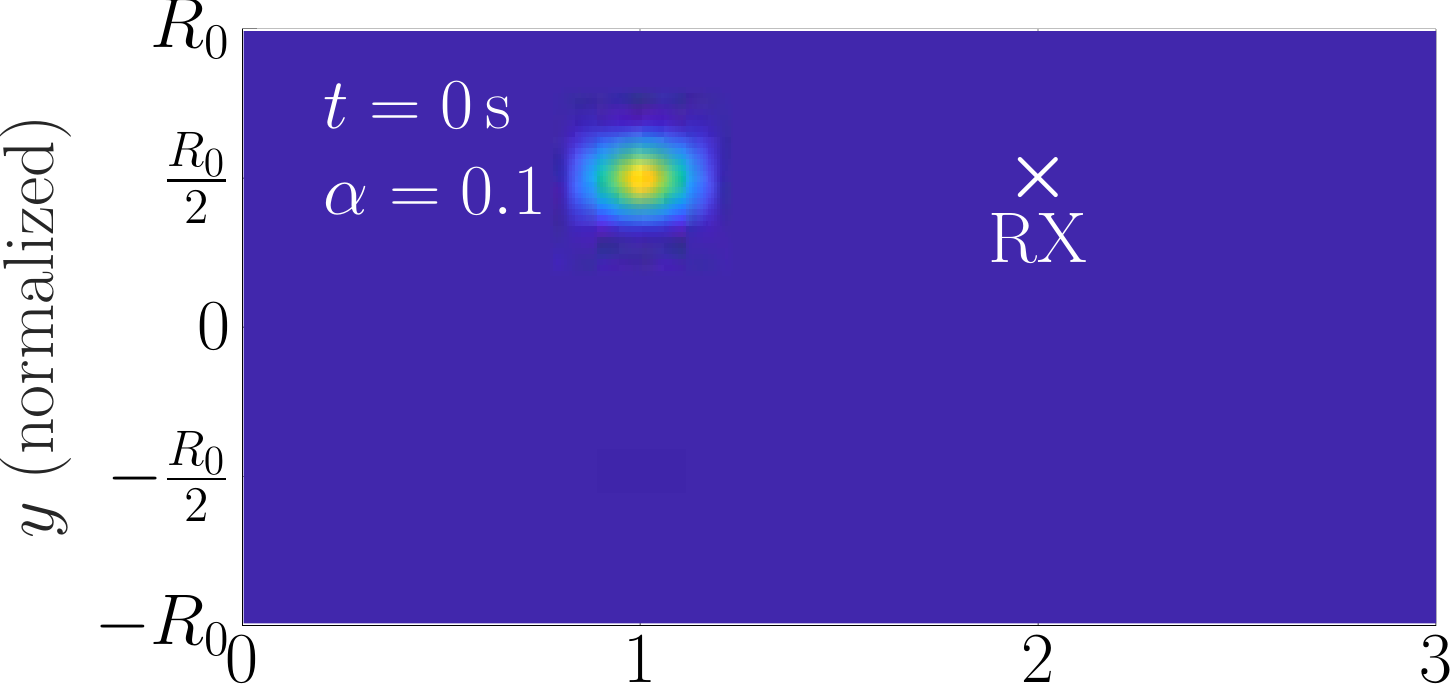}
    \end{minipage}\hfill
    \begin{minipage}{0.3\textwidth}
    	\centering
    	\vspace*{0.5ex}	
	\includegraphics[width=0.85\linewidth]{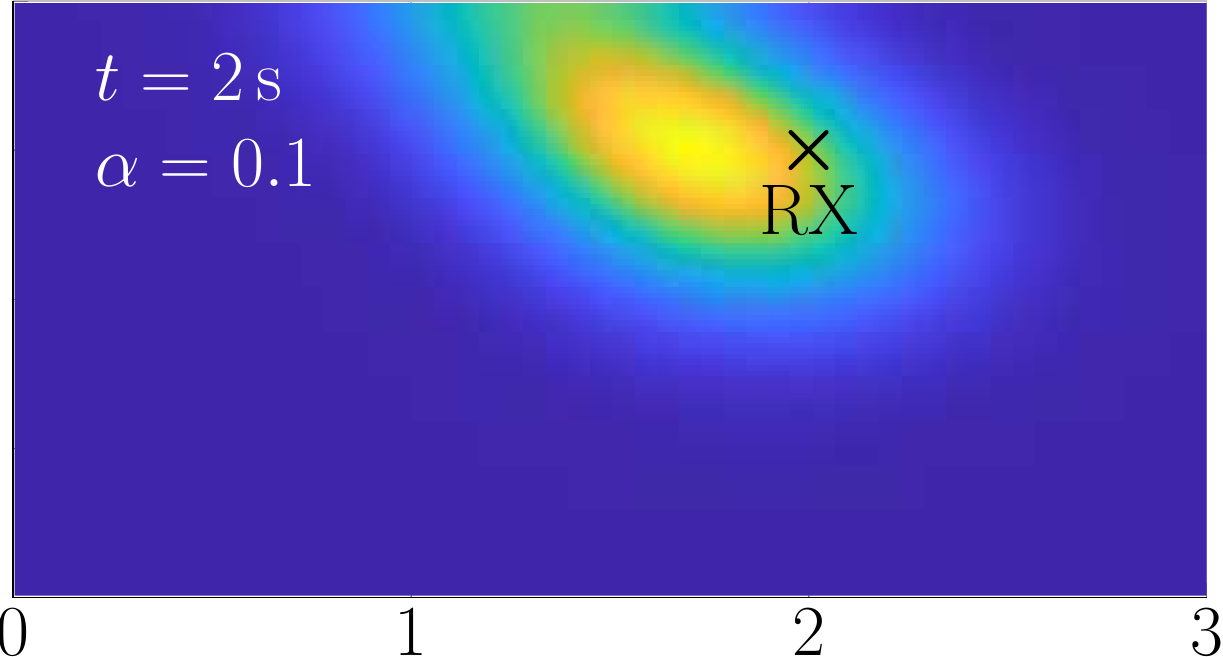}
    \end{minipage}\hfill
    \begin{minipage}{0.3\textwidth}
    	\centering
	    	\vspace*{0.5ex}
	\includegraphics[width=0.85\linewidth]{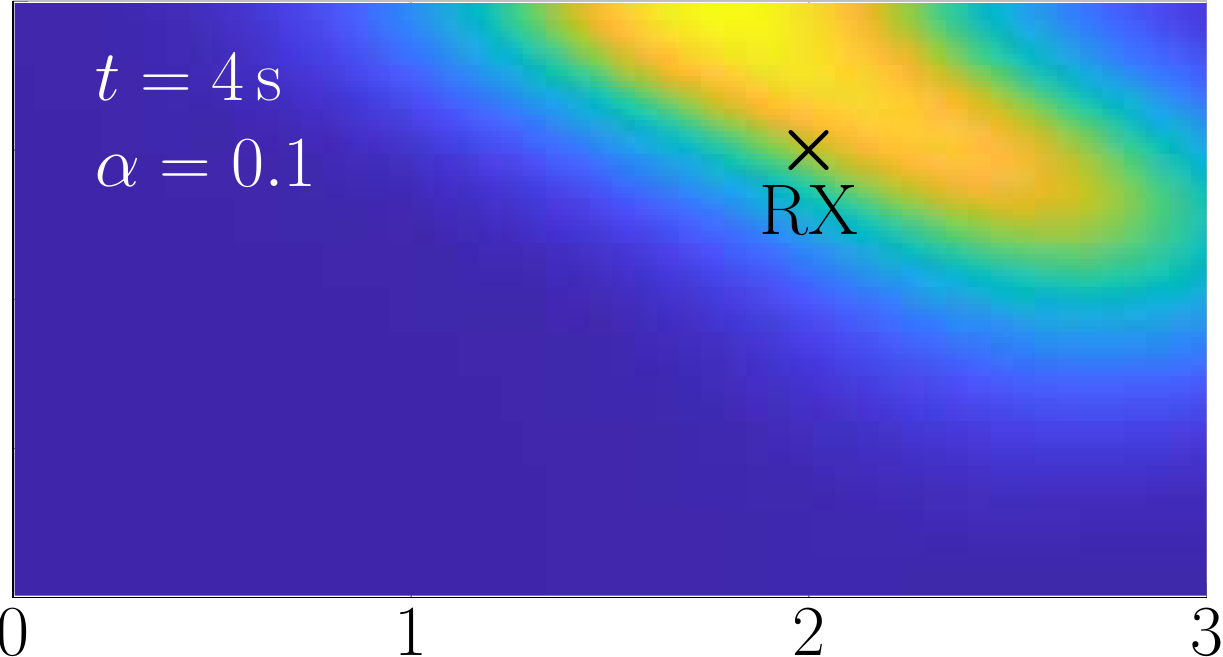}
    \end{minipage}\\
    \begin{minipage}{0.3\textwidth}
    	\centering
        \includegraphics[width=\linewidth]{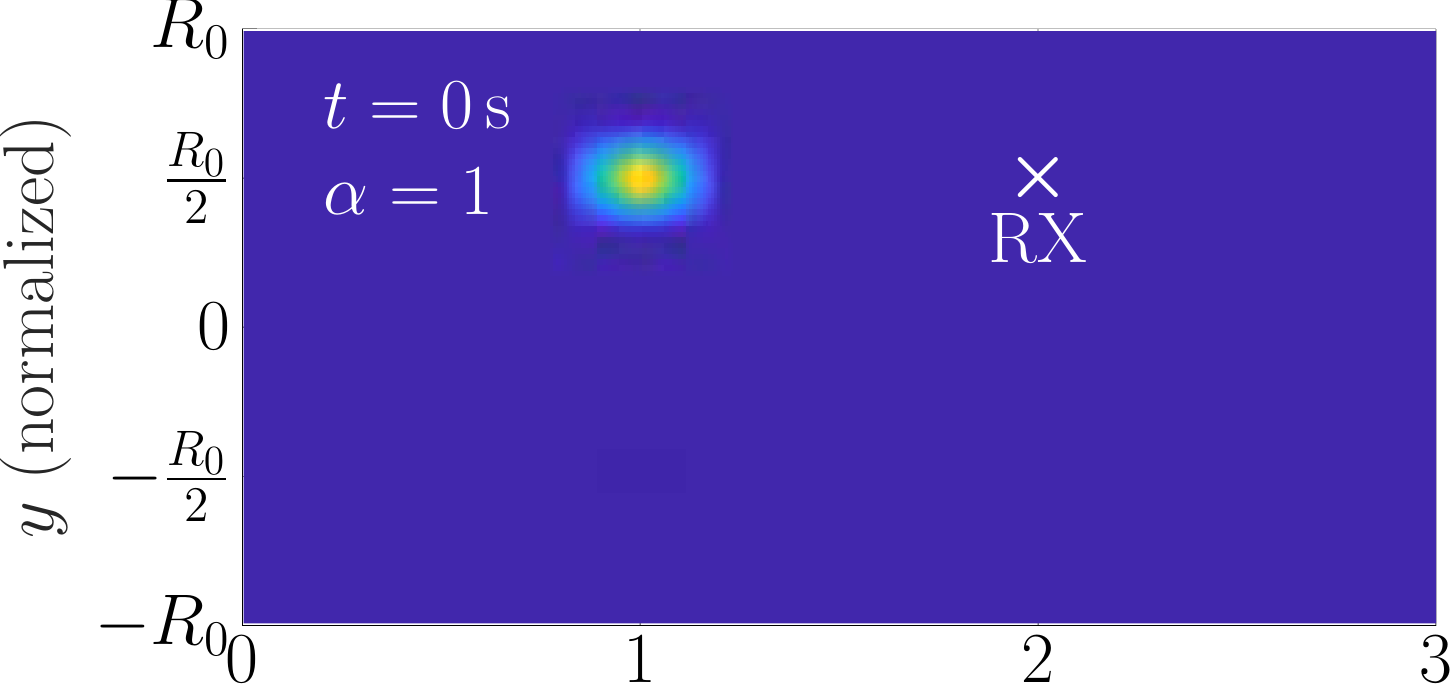}
    \end{minipage}\hfill
    \begin{minipage}{0.3\textwidth}
    	\centering
    		\hspace*{0.25ex}
	    	\vspace*{-0.5ex}
	\includegraphics[width=0.85\linewidth]{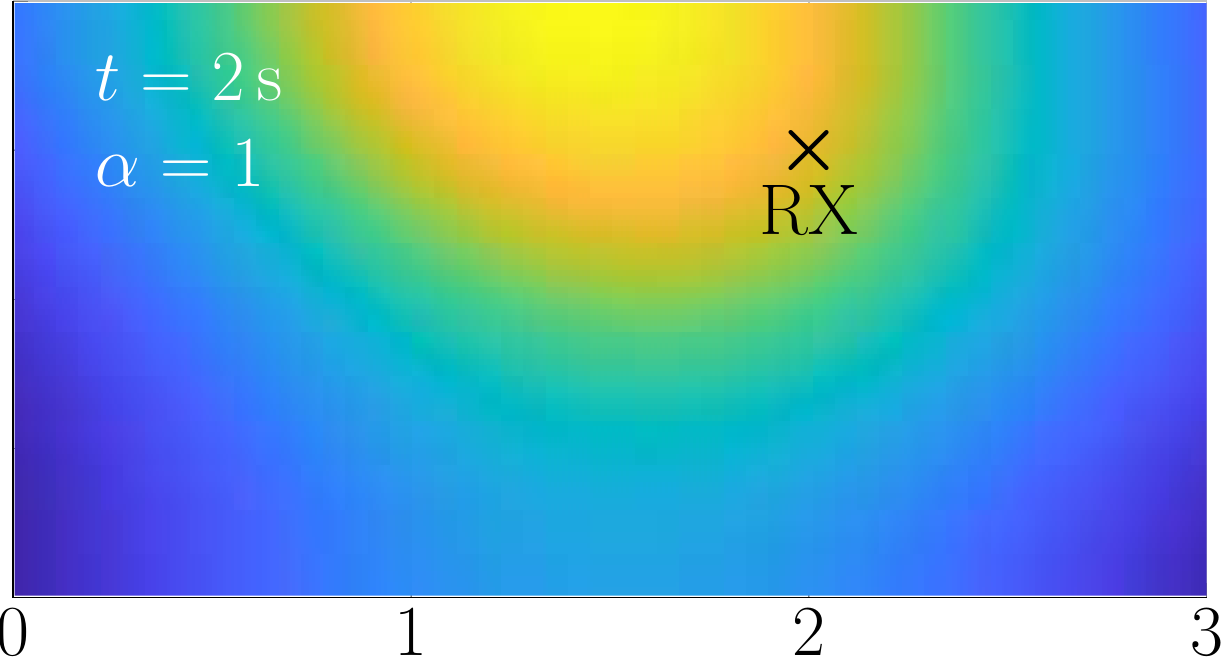}
    \end{minipage}\hfill
    \begin{minipage}{0.3\textwidth}
    	\centering
    	    \hspace*{0.25ex}
	    	\vspace*{-0.5ex}
	\includegraphics[width=0.85\linewidth]{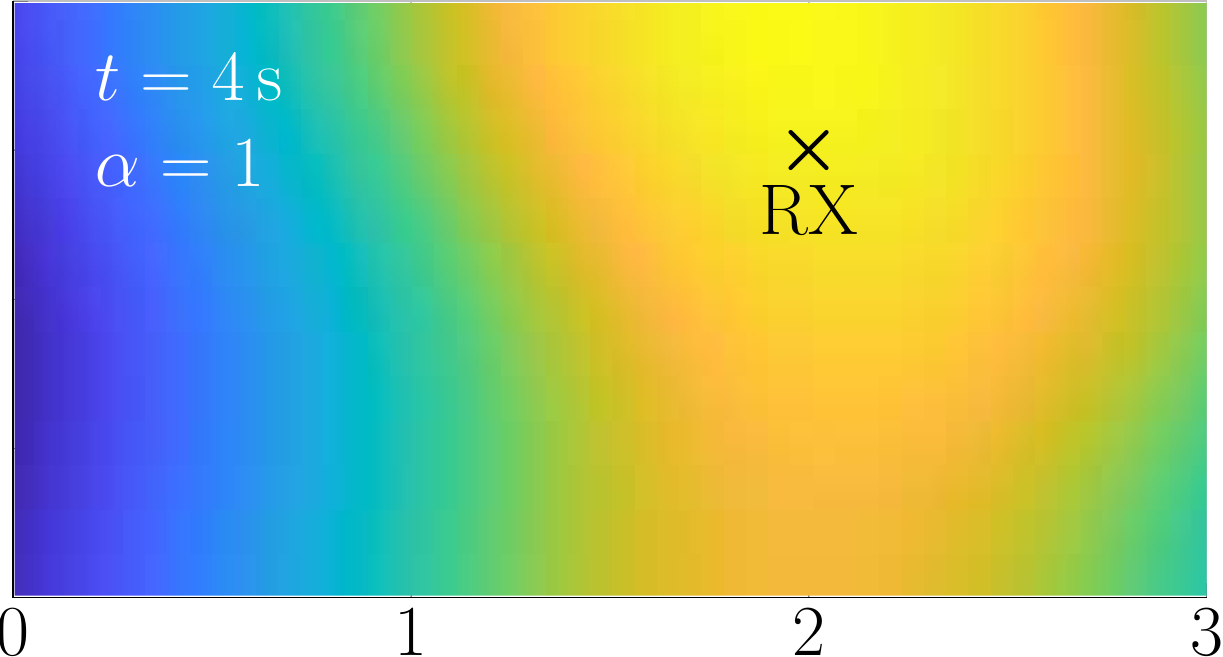}
    \end{minipage}
    \vspace{-1ex}
    \caption{\small 2D concentration $p(\bm{x},t)$ in the $y$-$z$-plane ($y = r\,\sin(\frac{\pi}{2})$) of the cylinder at times $t = 0\,\si{\second},  2\,\si{\second},  4\,\si{\second}$ for a point release.
        Different values $\alpha = 1\cdot 10^{-3},\, 0.1,\, 1$ are employed (top to bottom). 
        }
    \label{fig:point2D}
    \vspace*{-3.5ex}
\end{figure*}

Fig.~\ref{fig:point2D} shows the concentration for a point release at $r_\mathrm{e} = 0.5$ in the cylinder for different values of $\alpha$.
Although the general propagation behavior for a point release is similar to that for a uniform release (see Fig.~\ref{fig:uniform2D}), Fig.~\ref{fig:point2D} highlights some differences and reveals significant effects of practical relevance. 
The top row of Fig.~\ref{fig:point2D} shows the propagation for $\alpha = 1\cdot 10^{-3}$. As previously discussed, flow dominates diffusion in this case and the initial point is not spread spatially, but its initial shape is still distorted over time due to the parabolic flow profile, see top row of Fig.~\ref{fig:uniform2D}. 
For $\alpha = 0.1$ the point starts to spread by diffusion, see middle row of Fig.~\ref{fig:point2D}. Due to the radial release position at $r_\mathrm{e} = 0.5$, and the the zero flow $v(R_0) = 0$ at the boundary, a certain percentage of particles accumulate at the cylinder wall for $t = 4\,\si{\second}$ where they are only affected by diffusion. 
Furthermore, particles start to propagate into the lower part of the cylinder. This effect is even more pronounced for $\alpha = 1$, see bottom row of Fig.~\ref{fig:point2D}. In this case, after $t = 4\,\si{\second}$, the particles are distributed over the entire radial plane of the cylinder.
The effects that arise for increasing $\alpha$ and $t$ raise the question for which $\alpha$ the initial TX position is completely forgotten for a given RX position.

\begin{figure*}[t]
    \centering
    \begin{subfigure}[b]{0.3\textwidth}
            \centering
            	\includegraphics[width=0.9\linewidth]{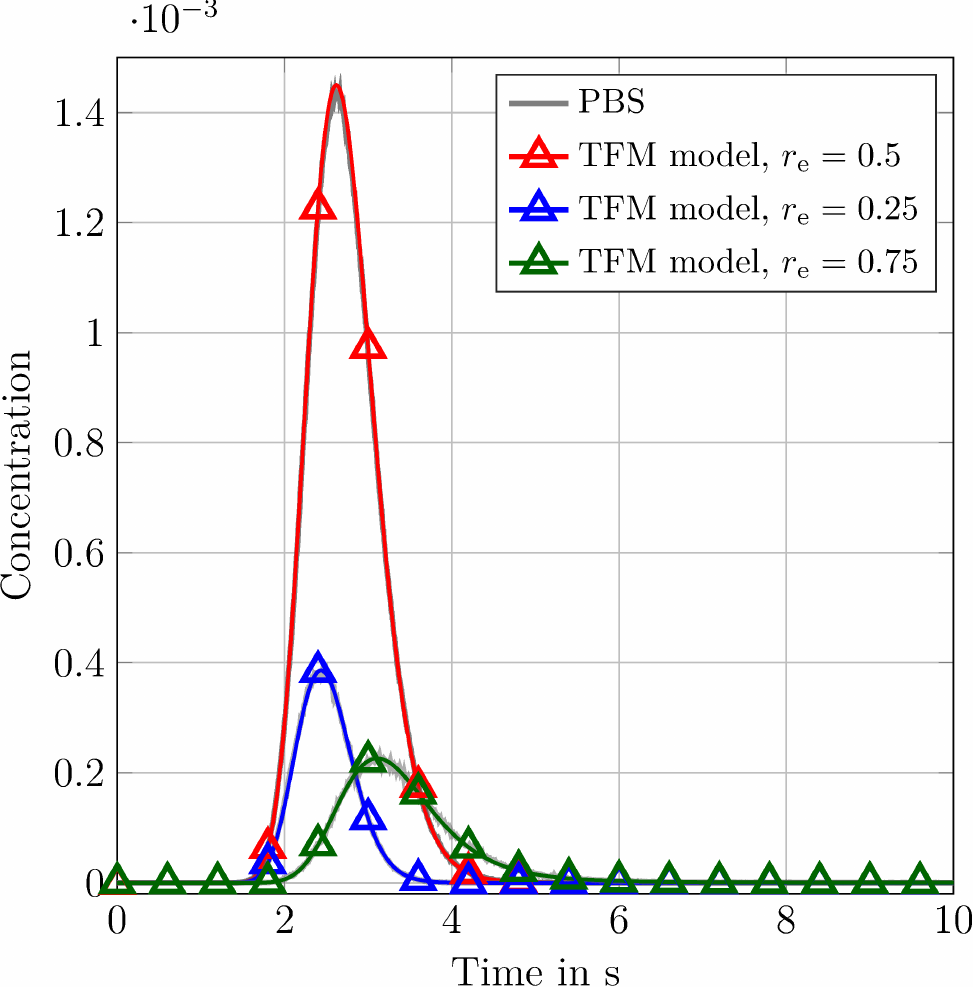}
            	\vspace{-1ex}
            \caption{$\alpha = 1\cdot 10^{-2}$}
    \label{fig:point1D:a}
    \end{subfigure}
    \begin{subfigure}[b]{0.3\textwidth}
            \centering
            \includegraphics[width=0.9\linewidth]{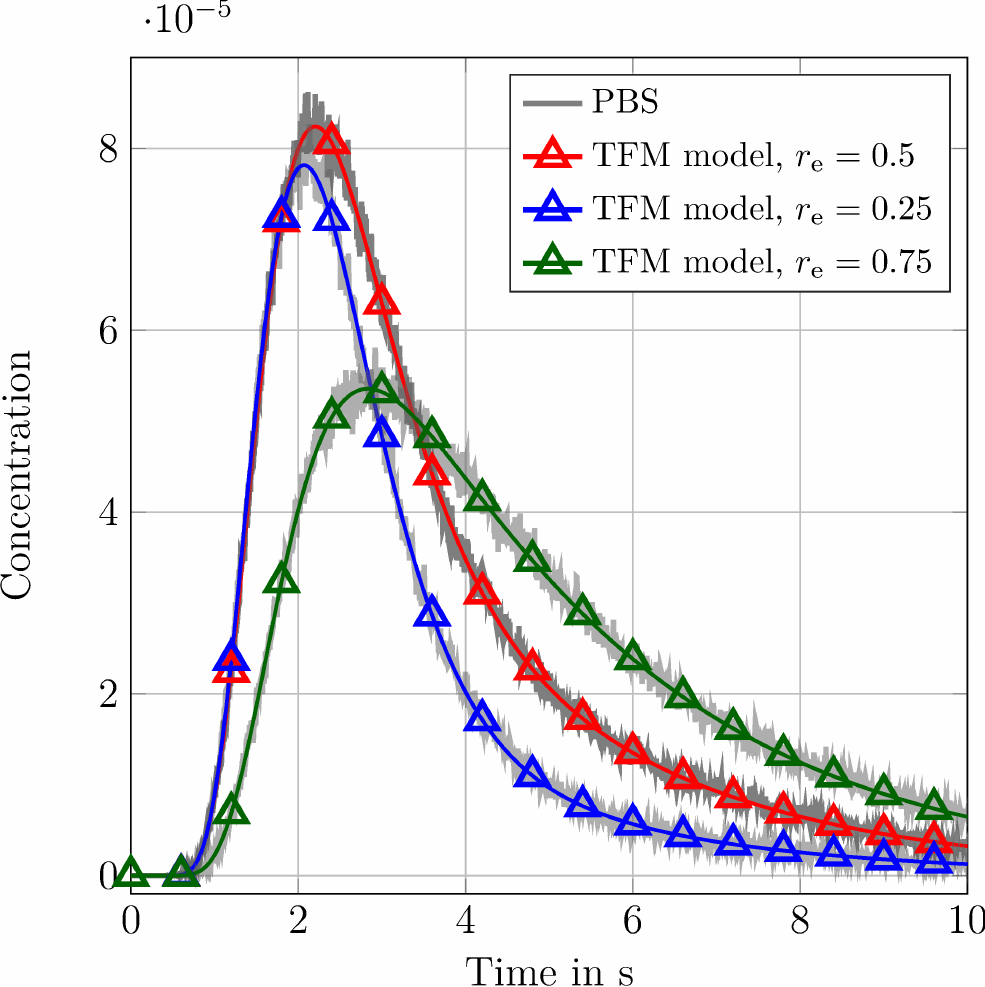}
            	\vspace{-1ex}
            \caption{$\alpha = 0.1$}
    \label{fig:point1D:b}
    \end{subfigure}    
    \begin{subfigure}[b]{0.3\textwidth}
            \centering
            \includegraphics[width=0.9\linewidth]{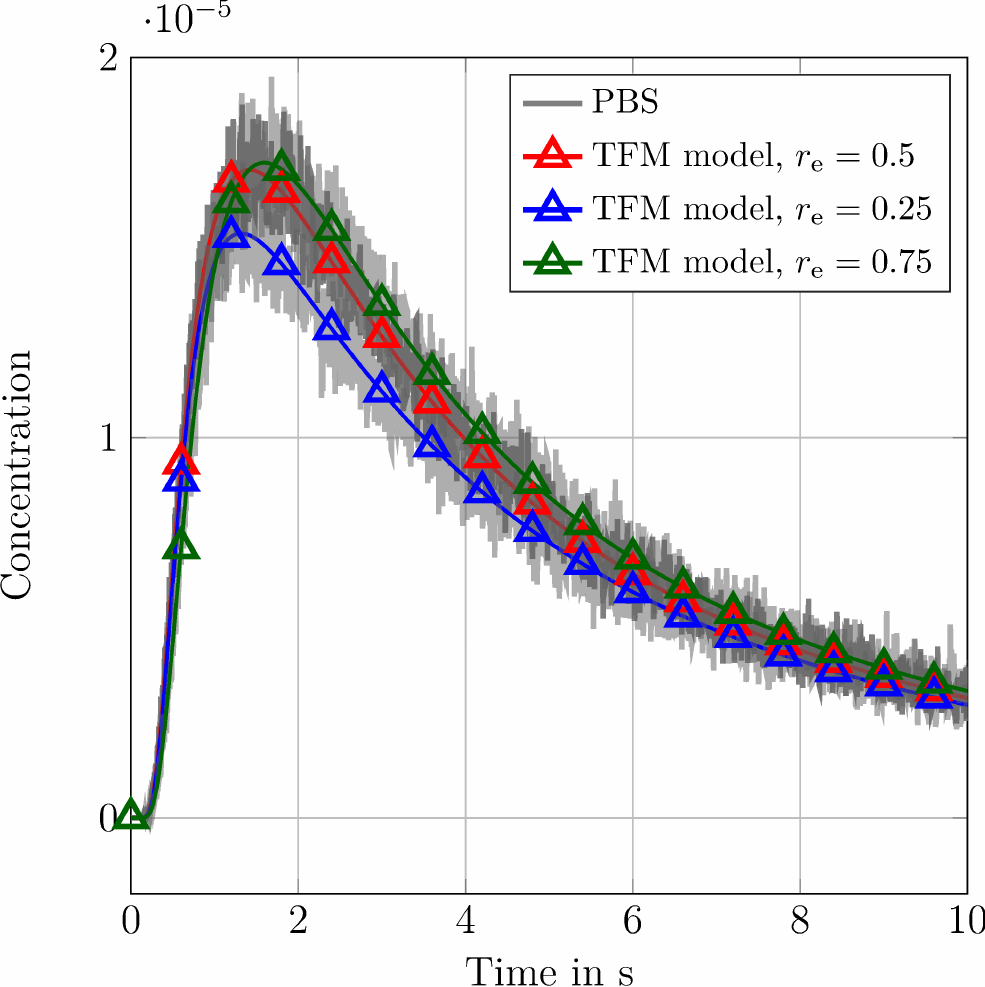}
            	\vspace{-1ex}
            \caption{$\alpha = 0.5$}
    \label{fig:point1D:c}
    \end{subfigure}
    \vspace{-2ex}
    \caption{\small 
    Concentration $p_\mathrm{cube}$ at $\bm{x}_\mathrm{RX}$ over time for a point release of particles at radial release positions $r_\mathrm{e} = 0.25,\, 0.5,\, 0.75$. Different values of $\alpha$ are considered.
    }
    \label{fig:point1D}
    \vspace*{-3.8ex}
\end{figure*}

This question is further investigated in Fig.~\ref{fig:point1D}, which shows the concentration \eqref{eq:70} at position $\bm{x}_\mathrm{RX}$ for different values of $\alpha$ after a point release on the radial axis for $r_\mathrm{e} = 0.25,\, 0.5$  and $0.75$. 
The figure shows the numerical evaluation of the proposed model (red, blue, green colors) and for comparison results from PBS (gray color). 
For $\alpha = 1\cdot 10^{-2}$ (flow dominant regime,  Fig.~\ref{fig:point1D:a}), the differences in the received concentration caused by different release positions are clearly visible.
For $\alpha = 0.1$ (mixed regime,  Fig.~\ref{fig:point1D:b}) the received concentration starts to increase simultaneously for $r_\mathrm{e} = 0.5$ and $r_\mathrm{e} = 0.25$, while the received concentration starts to increase later for $r_\mathrm{e} = 0.75$, which is due to the zero flow at the boundary.
For $\alpha = 0.5$ (mixed regime, Fig.~\ref{fig:point1D:c}) the received concentration increases simultaneously for all considered release positions $r_\mathrm{e}$. 
In this scenario, the previously mentioned effect becomes evident, i.e., it is not possible to determine the release position $r_\mathrm{e}$ based on the received concentrations. 
This effect becomes even more pronounced for larger $\alpha$, see supplementary material in \cite[Sec.~3]{supplementary}.
We note that the numerical results obtained with the proposed TFM perfectly match the PBS results for all scenarios considered in Fig.~\ref{fig:point1D}.


\section{Implementation and Analysis}
\label{sec:analysis}
The previous section has shown that the evaluation of the proposed model perfectly matches the results obtained with PBS for all considered scenarios. 
The proposed TFM is an analytical solution for the advection-diffusion problem and provides a compact description valid for all regimes. Therefore, the model closes the gap between existing solutions for the flow dominant and dispersive regimes.  

In this section, we provide a short overview of the implementation of the proposed discrete-time SSD \eqref{eq:60}, \eqref{eq:61}. Furthermore, the accuracy of the model is analyzed and its limitations are discussed. Finally, the benefits of the proposed model are shortly summarized.

\subsection{Remarks on the Implementation}
\label{subsec:anal:implementation}
\vspace*{-1ex}
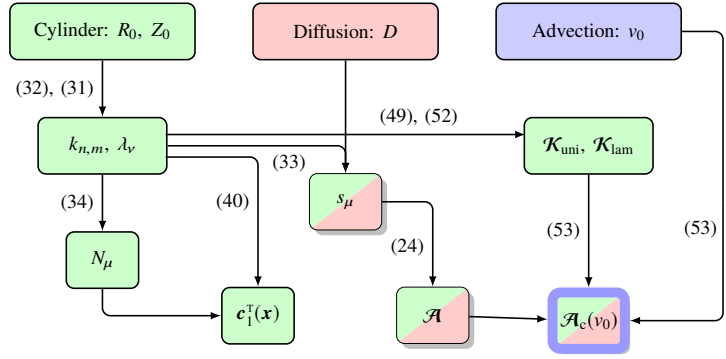
\begin{figure*}[t]
	\centering
	\scalebox{0.75}{
	\begin{tikzpicture}[
	box/.style={rectangle,draw,thick,rounded corners, text width=2cm,minimum height=10mm, node distance=1mm, align=center},
	arrow/.style={-latex,thick,draw,rounded corners}
	]
	\node[box, text width=3cm, fill=green!20] (geometry) at (0,0) {\small Cylinder: $R_0$, $Z_0$};
	\node[box, fill=red!20] (diffusion) [right = 10mm of geometry, text width=3cm] {\small Diffusion: $D$};
	\node[box, fill=blue!20] (flow) [right = 10mm of diffusion, text width=3cm] {\small Advection: $v_0$};
	
	\node[box] (k_lam) [below = 10mm of geometry, text width=2cm, fill=green!20] {\small $k_{n,m}$, $\lambda_\nu$};
	\node[diagonal fill={red!20}{green!20},text centered, rounded corners, draw, drop shadow] (smu) [below = 20mm of diffusion, text width=1cm,minimum height=10mm] {\small $s_\mu$};
	\node[box] (klam) [below = 10mm of flow, text width=2cm, fill=green!20] {\small $\Ks_{\mathrm{uni}}$, $\Ks_{\mathrm{lam}}$};	
	\node[box] (nmu) [below = 10mm of k_lam, text width=1cm, fill=green!20] {\small $N_\mu$};
	\node[box] (cx) [below left = 10mm and 2.5mm of smu, text width=1cm, fill=green!20] {\small $\bm{c}_1\tran(\bm{x})$};	
	\node[diagonal fill={red!20}{green!20},text centered, rounded corners, draw, drop shadow] (smu) [below = 20mm of diffusion, text width=1cm,minimum height=10mm] (as) [below right = 10mm and 2.5mm of smu, text width=1cm] {\small $\As$};	
	
	\node[diagonal fill={red!20}{green!20},text centered, rounded corners, draw, drop shadow] (smu) [below = 20mm of diffusion, text width=1cm,minimum height=10mm, draw = blue!40, line width = 5pt] (ac) [below = 20mm of klam, text width=1cm] {\small $\As_\mathrm{c}(v_0)$};
		
	\draw[arrow](geometry) -- node[midway,left]{\small\eqref{eq:30}, \eqref{eq:31}} (k_lam);
	\draw[arrow](k_lam) -- node[midway,left]{\small\eqref{eq:33}} (nmu);
	\draw[arrow](k_lam) -| node[midway, below, xshift=-10mm]{\small\eqref{eq:32}} (smu);
	\draw[arrow](nmu) |- (cx);
	
	\draw[arrow](diffusion) -- (smu);
	\draw[arrow]($(k_lam) - (-1.12cm,2mm)$) -| node[midway, left, yshift=-8mm]{\small\eqref{eq:39}} (cx);
	\draw[arrow](smu) -| node[midway, left, yshift=-8mm]{\small\eqref{eq:23}} (as);
	
	\draw[thick, rounded corners](flow) -| node[midway,left,yshift = -35mm, xshift=0.5mm]{\small\eqref{eq:49}}($(k_lam) + (11cm,-1cm)$);
	\draw[arrow] ($(k_lam) + (11cm,-1cm)$)|- (ac); 
	\draw[arrow]($(k_lam) + (1.12cm,2mm)$) -- node[midway, above, xshift=13mm]{\small\eqref{eq:46a}, \eqref{eq:48a}} ($(klam) + (-1.12cm,2mm)$);
	
	\draw[arrow](as) -- (ac);
	\draw[arrow](klam) -- node[midway,left]{\small\eqref{eq:49}}(ac);
	\end{tikzpicture}
	}
	\vspace{-2ex}
	\caption{\small 
	Calculation of vectors and matrices in the SSD \eqref{eq:60}, \eqref{eq:61} and their dependencies on physical parameters $D$, $v_0$, and geometrical parameters $R_0$, $Z_0$.}
	\label{fig:dependencies}
	\vspace*{-3.8ex}
\end{figure*}

To analyze the dynamics of the particles in the cylinder, the derived discrete-time model in \eqref{eq:60}, \eqref{eq:61} has to be numerically evaluated. 
%
The dependence of variables $\bm{c}_1$ and $\As_\mathrm{c}$ in \eqref{eq:60}, \eqref{eq:61} on the physical and geometrical parameters is illustrated in Fig.~\ref{fig:dependencies}. In the figure, green boxes indicate a dependence on geometrical parameters $R_0$, $Z_0$, red boxes a dependence on the diffusion coefficient $D$, and blue boxes a dependence on the flow velocity $v_0$.  
As can be observed, the values $k_{n,m}$ and $\lambda_\nu$ only depend on the geometry of the cylinder and can be computed independent from diffusion coefficient $D$ and flow velocity $v_0$. The same applies for both  
feedback matrices $\Ks$.
By exploiting these limited dependencies, many calculations needed for the evaluation of the SSD \eqref{eq:60}, \eqref{eq:61} can be performed once in advance and do not have to be repeated if the parameters change.  
%

Depending on the number of eigenvalues $Q$, see \eqref{eq:65}, a straightforward implementation of the SSD \eqref{eq:60}, \eqref{eq:61} may lead to high computational costs. 
Particularly, the calculation and subsequent multiplication of the potentially fully occupied matrix $\As_\mathrm{c}^\mathrm{d}$ and system states $\bar{\bm{y}}^\mathrm{d}$ in \eqref{eq:60} is time consuming.
Therefore, state equation \eqref{eq:60} should be modified to speed up the required multiplications. 
To this end, the block diagonal structure of matrix $\As_\mathrm{c}^\mathrm{d}$ can be exploited, i.e., the matrix consists of $(2N +1)$ blocks of size $(M\cdot L) \times (M\cdot L)$. This block structure allows the block-wise calculation of \eqref{eq:60}.
%
%
Each of the resulting blocks can be further simplified by an eigendecomposition which allows for fast evaluation by a parallel structure of filters. Furthermore, due to the symmetry of the Bessel functions, the number of blocks can be reduced to $(N+1)$, see Section~\ref{subsec:sim:point}. 
These modifications can be applied to enable fast evaluation of the proposed model in, e.g., MATLAB. They are further described in the supplementary material provided together with the MATLAB code in \cite{supplementary}.

\subsection{Analysis of Accuracy}
\label{subsec:accuracy}
\vspace*{-1ex}
Although the proposed model involves an infinite sum, see \eqref{eq:25}, it represents an analytical solution for the considered advection-diffusion process. 
Mathematically, the solution only converges if the number of eigenvalues $Q \to \infty$. 
For numerical evaluation and analysis, this number has to be restricted to a finite value. 
This implies a trade-off between the complexity and accuracy of the model, see also \cite[Section~IV-C]{schaefer:icc:2019}. 
In practice, the number of eigenvalues $Q$ has to be chosen such that the accuracy requirements of the desired scenario are met.
For example, if only a rough impression of the system behavior is desired, a low number $Q$ would be sufficient. For an accurate evaluation of the concentration in the complete volume, a higher value of $Q$ is needed.

\begin{figure*}[t]
	\centering
    \begin{subfigure}[b]{0.3\textwidth}
            \centering
            	\includegraphics[width=0.9\linewidth]{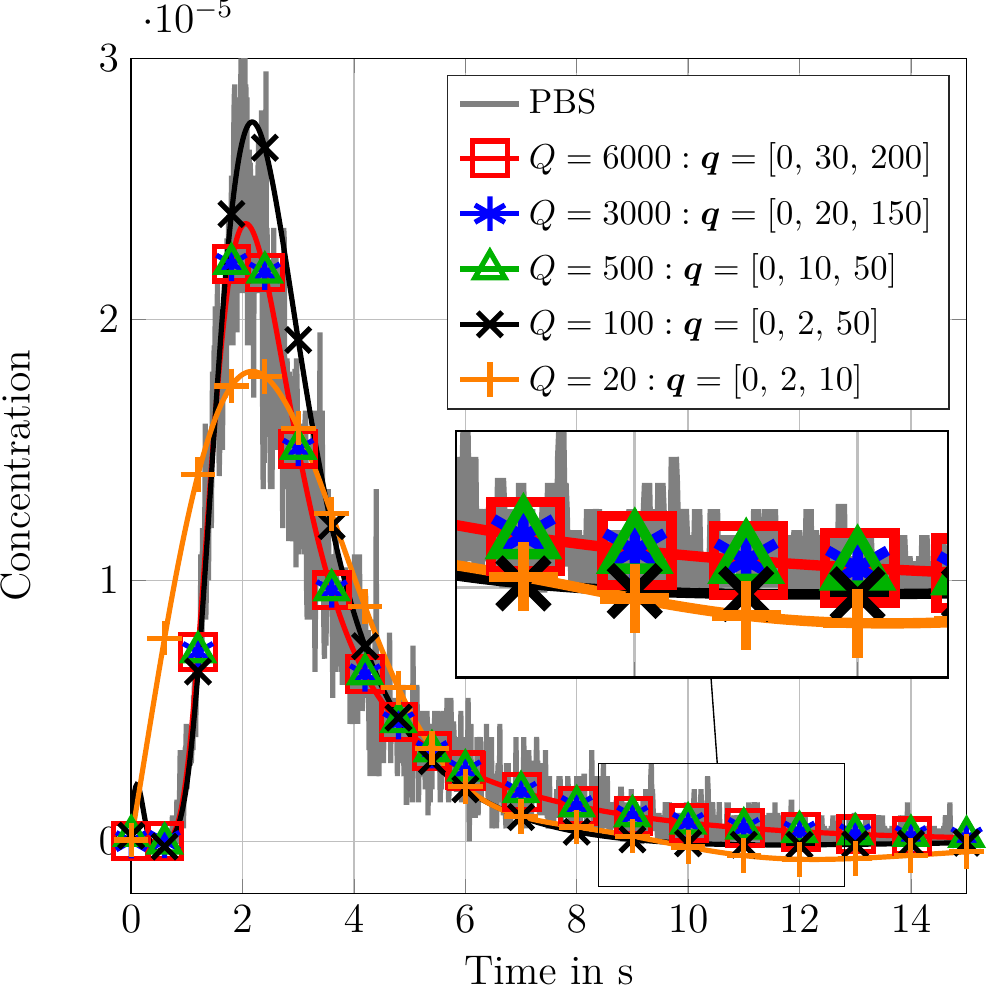}
            	\vspace{-1ex}	
            \caption{$\alpha = 0.1$, uniform release}
    \label{fig:accuracy:a}
    \end{subfigure}
    \begin{subfigure}[b]{0.3\textwidth}
            \centering
            \includegraphics[width=0.9\linewidth]{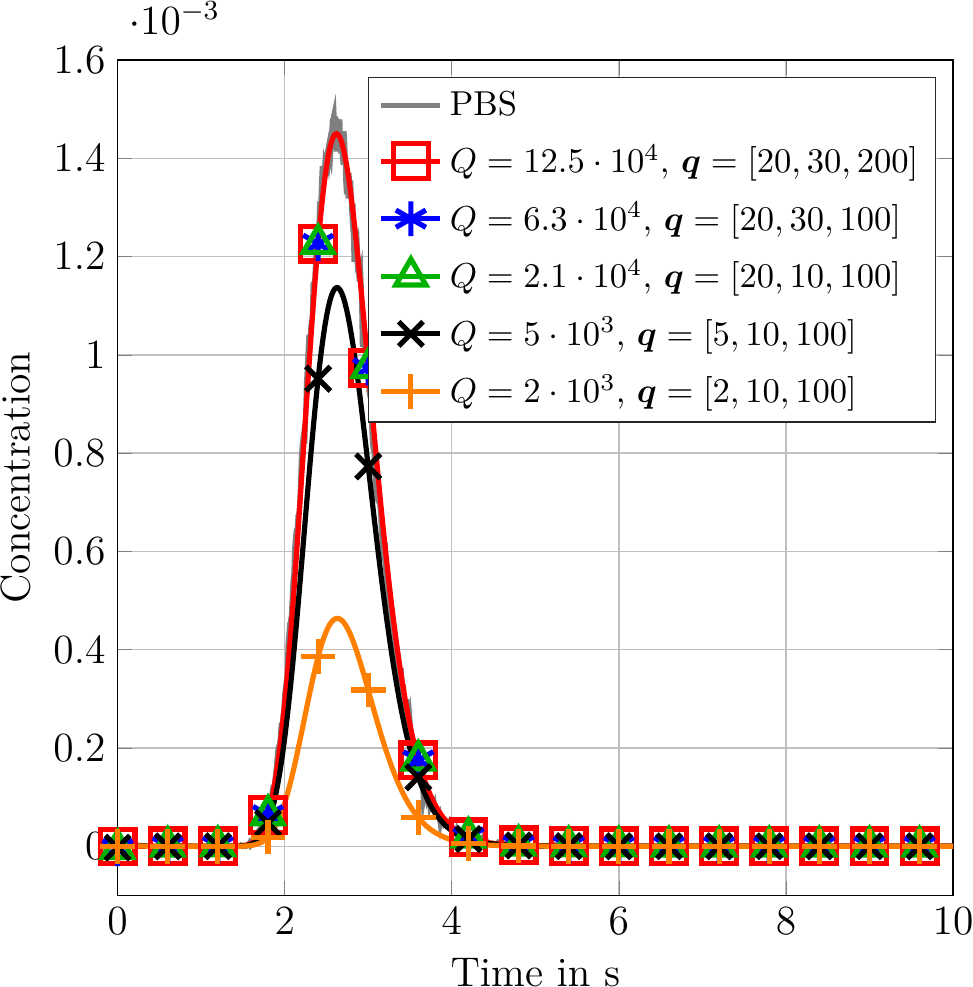}
            	\vspace{-1ex}
            \caption{$\alpha = 1\cdot 10^{-2}$, point release}
    \label{fig:accuracy:b}
    \end{subfigure}
    \begin{subfigure}[b]{0.3\textwidth}
            \centering
            	\includegraphics[width=0.9\linewidth]{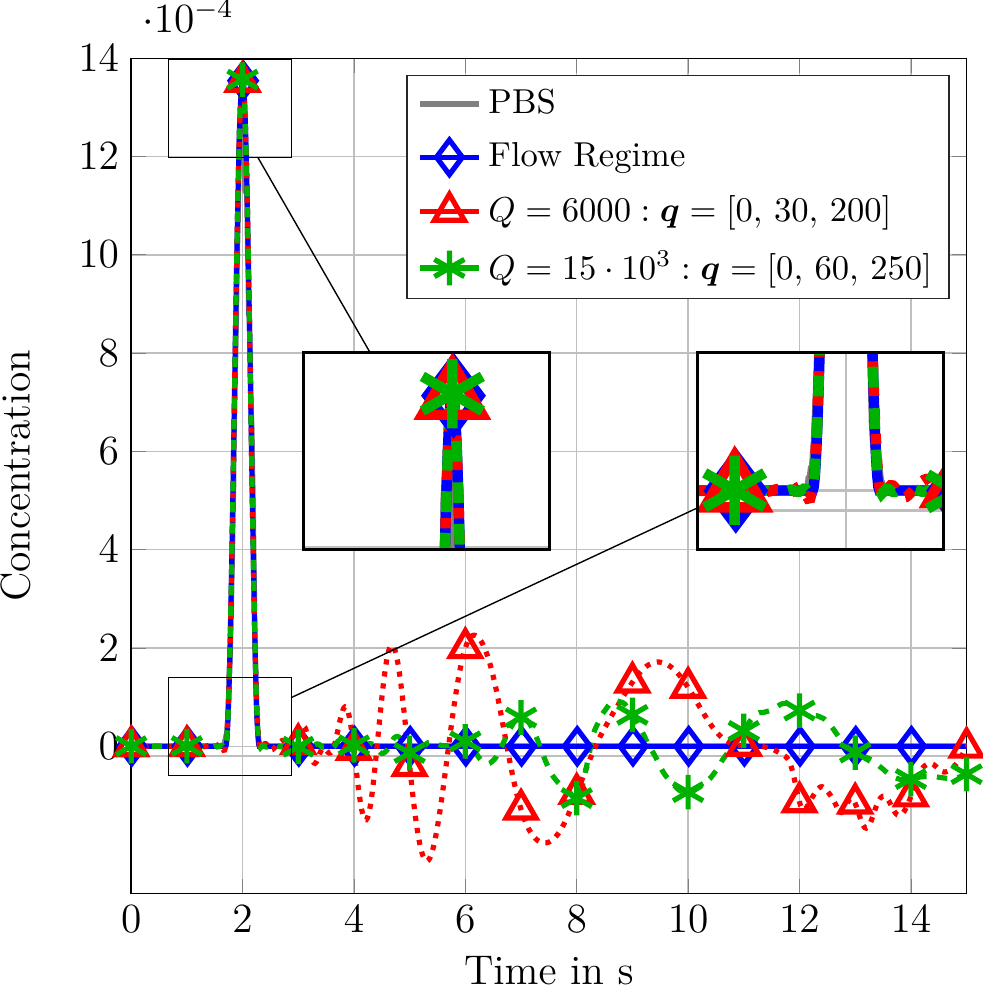}
            	\vspace{-1ex}
            \caption{$\alpha = 1\cdot 10^{-5}$, uniform release}
    \label{fig:accuracy:c}
    \end{subfigure}
    \vspace{-2ex}
	\caption{\small Accuracy analysis for (a) uniform release with $\alpha = 0.1$, (b) point release width $\alpha = 1\cdot 10^{-2}$, and (c) uniform release with $\alpha = 1\cdot 10^{-5}$.}
	\label{fig:accuracy}
	\vspace{-3.8ex}
\end{figure*}

In Figs.~\ref{fig:accuracy}\subref{fig:accuracy:a} and \subref{fig:accuracy:b}, the proposed model is evaluated for different values of $Q$ and PBS results are provided as ground truth. 
The scenarios considered in Figs.~\ref{fig:accuracy}\subref{fig:accuracy:a} and \subref{fig:accuracy:b} are identical to those in Figs.~\ref{fig:uniform1D:c} and \ref{fig:point1D:a} ($r_\mathrm{e} = 0.5$). The accuracy for uniform release and point release are analyzed separately because for the uniform release only Bessel functions of order $n = 0$ contribute to the solution. 
For uniform release, the proposed model converges to the PBS result for $Q = 6000$. 
Reducing the number of eigenvalues to $Q = 3000$ or even $Q = 500$, the proposed model is still in good agreement with the PBS results. Only for $Q = 100$ and $Q = 20$ a significant difference can be observed.  
In the point release case, a higher number of eigenvalues is necessary for convergence. 
Fig.~\ref{fig:accuracy:b} shows that the proposed model converges to the PBS results for $Q = 2.1\cdot 10^{4}, \dots, 12.5\cdot 10^{4}$ eigenvalues. 
For smaller values $Q = 5\cdot 10^{3}$ and $Q = 2\cdot 10^{3}$, the proposed model is not converging to the PBS results. 
Compared to a uniform release, larger values of $Q$ are necessary for convergence for a point release because also Bessel functions of order $n\neq 0$ contribute to the solution. Particularly, large values of $N$ are necessary to correctly represent the particle propagation in the volume (see videos for small values of $N$ in \cite[Sec.~4.3]{supplementary}). 

Fig.~\ref{fig:accuracy:c} illustrates a limitation of the proposed model, i.e., of its numerical evaluation, arising for very low values of $\alpha$, e.g., $\alpha = 1\cdot 10^{-5}$. 
The figure shows the numerical evaluation of the proposed model for different values of $Q$ while PBS results and the known solution for the flow dominant regime are provided as ground truth. 
From $t = 0\,\si{\second}$ up to $t \approx 2.5\,\si{\second}$, it can be observed that the evaluation of the proposed model still perfectly matches the PBS results and the known solution for the flow dominant regime. 
Particularly, the amplitude, the duration, and the shape of the received concentration from a uniform release are captured by the proposed model, see the zoomed excerpts in Fig.~\ref{fig:accuracy:c}, while undesired oscillations occur after the peak. 
As explained in Section~\ref{sec:tfm2}, in the proposed model, the influence of flow is incorporated by a shift of the eigenvalues, the accuracy of which depends on the number of eigenvalues $Q$. 
However, for very small values of $\alpha$, the influence of flow dominates and therefore, the shift of the eigenvalues is very large and the proposed model is not converging for the considered values of $Q$. 
By increasing the number of eigenvalues significantly to $Q = 15\cdot 10^{3}$ (green curve in Fig.~\ref{fig:accuracy:c}), the amplitude of the oscillations can be reduced but the effect cannot be suppressed completely.  
The analytical form of the proposed model, e.g., in terms of a CGF in \eqref{eq:58}, is not necessarily restricted to a finite number of eigenvalues $Q$. 
Therefore, the limitation for very small values of $\alpha$ does not affect the validity of the proposed analytical model, but only its numerical evaluation where $Q$ has to be finite. 
Hence, for very small values of $\alpha$, using the known solution for the flow dominant regime may be preferable, see, e.g., \cite[Eq.~(16)]{wicke:globecom:2018}.
As a rule of thumb, the number of eigenvalues $Q$ required for an accurate representation decreases for increasing values of $\alpha$ and for increasing spatial spreading of the initial particle distribution. 
This is because fewer spatial eigenfunctions are needed to approximate smooth functions compared to peaky ones. 

In Section~\ref{sec:math}, the cylinder is bounded in $z$-direction by BCs \eqref{eq:3} and \eqref{eq:9}. 
Particularly, BC \eqref{eq:9} corresponds to an absorbing boundary, and therefore all particles leave the cylinder for $t\to\infty$. 
This is true for the analytical formulation of the proposed model, but for its numerical evaluation some undesired effects occur due to the numerical restriction of the $z$-direction to $Z_0$, see video in \cite[Sec.~4.1]{supplementary}. 
Instead of leaving the cylinder at $z = Z_0$, particles are reflected and re-enter the cylinder at $z = 0$ where the re-entering is accompanied by undesired reflections in the cylinder. 
In future work, we plan to overcome this effect by adopting the techniques proposed in \cite{Grant:absbound:2015}. 
For numerical evaluation of the model in the proposed form, these effects can be avoided, e.g., by ensuring that no particles leave the cylinder during the observation time of the system. Therefore, we restricted the observation time to $t_\mathrm{obs} \leq \frac{Z_0 - z_\mathrm{e}}{v_0} =18\,\si{\second}$ for the numerical evaluation in the considered scenarios.

%

%

\begin{table}[t]
\caption{\small Different formulations of the proposed model.}
\label{tab:tfm}
\vspace{-1ex}
\centering
\begin{tabular}{p{0.5cm}|p{8cm}|p{1.2cm}|p{3.5cm}}
\hline\noalign{\smallskip}
PDE & 
$\frac{\partial}{\partial t}p(\bm{x},t) = D\,\mbox{div}\left(\mbox{grad}\left( p(\bm{x},t)\right)\right) - v(r)\,\mbox{div} \left(p(\bm{x},t) \bm{e}_z(\bm{x})\right)$
& Eq.~\eqref{eq:4},\eqref{eq:5} & continuous-time domain\\[3pt]
\hline\noalign{\smallskip}
CGF & 
$p(\bm{x},t) = \int_{V} g(t, \bm{x}\vert \bm{\xi}) p_\mathrm{i}(\bm{\xi})\dint{\bm{\xi}}$
& Eq.~\eqref{eq:58} & continuous-time domain\\[3pt]
\hline\noalign{\smallskip}
TFM & 
$P(\bm{x},s) = \bm{c}_1\tran(\bm{x})\bar{\bm{H}}(s,D,v_0)\left[\bar{\bm{F}}_\mathrm{e}(s) + \bar{\bm{y}}_\mathrm{i} \right]$
& Eq.~\eqref{eq:60a} & frequency domain\\[3pt]
\hline\noalign{\smallskip}
SSD & 
$\bar{\bm{y}}^\mathrm{d}[k] = \As_\mathrm{c}^\mathrm{d}(v_0)\bar{\bm{y}}^\mathrm{d}[k - 1] 
+ \bar{\bm{f}}^\mathrm{d}_\mathrm{e}[k] + \bar{\bm{y}}_\mathrm{i}\delta[k]$\newline
$p^\mathrm{d}[\bm{x},k] = \bm{c}_1\tran(\bm{x})\bar{\bm{y}}^\mathrm{d}[k]$
& Eq.~\eqref{eq:60}\newline
Eq.~\eqref{eq:61} & discrete-time domain\\[3pt]
\hline\noalign{\smallskip}
\end{tabular}
\vspace*{-2ex}
\end{table}

\subsection{Benefits of the Proposed Model}
\vspace*{-1ex}
In Section~\ref{sec:simul}, the proposed model matches the PBS results for all considered scenarios. 
Despite the previously mentioned limitations for the numerical evaluation of the proposed model, it provides a general analytical description of the advection-diffusion process with laminar flow. 
%
Section~\ref{sec:tfm2} introduced different equivalent formulations of the proposed model which are summarized in Table~\ref{tab:tfm}. The CGF in \eqref{eq:58} and the representation in terms of transfer functions in \eqref{eq:60a} provide an analytical description of the MC channel and allow a representation of the channel response in analytical form for given TX models. The discrete-time SSD in \eqref{eq:60}, \eqref{eq:61} provides a suitable model for numerical evaluation. 
The dependence of the convergence on the number of eigenvalues $Q$ (see Section~\ref{subsec:accuracy}) also implies flexibility. 
Particularly, the value of $Q$ can be adjusted depending on the objective of the investigation. 
To get a rough impression of the channel behavior for a given input signal, a small $Q$ is sufficient to perform many simulations in a short time, while a large value of $Q$ can be chosen for accurate simulation of particle propagation. 
The initial formulation of the model in terms of an SSD (see \eqref{eq:39}, \eqref{eq:49}) can also be extended as has been discussed for similar models in \cite[Sec.~IV-D]{schaefer:icc:2019}. 
For example, one possibility is the incorporation of more complex or time-varying boundary conditions, which is described in detail in \cite{Schaefer2020, rabenstein:ijc:2017}, and has been applied to cylindrical and spherical MC systems in \cite{schaefer:icc:2019, schaefer:icc:2020}. 
Furthermore, the SSD allows the interconnection of multiple systems (see \cite[Sec.~V]{schaefer:icc:2020}), which is beneficial for the modeling of interconnected tube systems or cascades of blood vessels. 
Finally, SSD models can be exploited for derivation of parameter estimation algorithms based on Kalman filters \cite{schaefer:ecc:2019} to determine relevant system parameters from measurements.

\section{Conclusion}
\label{sec:conc}
In this paper, we have proposed an analytical model for advection-diffusion processes in cylindrical environments affected by laminar flow. 
The proposed model has been derived based on a transfer function approach, which provides a general, flexible, and extendable description of the MC channel, and can be  formulated in terms of a CGF or an SSD depending on the application. 
The validity of the proposed model has been verified by numerical evaluation.
Particularly, it has been shown that the proposed solution matches the PBS in all considered regimes and corresponds to the known solutions for the flow dominant and dispersive regimes. 
In contrast to all known models, the proposed solution is also applicable in the mixed regime where both diffusion and flow have a similar impact on particle propagation.




%

The discussion of the benefits and limitations of the proposed model in Section~\ref{sec:analysis} suggests the following topics for future work: 
As discussed in Section~\ref{sec:analysis}, a large number of eigenvalues $Q$ may be necessary to fully capture all effects of the propagation of the particles in the cylinder. The required value of $Q$ can be reduced by a model reduction to the most dominant eigenvalues. 
Furthermore, to make the model even more comprehensive, the extension of the SSD  to more complex boundary conditions, e.g., semi-permeable walls, and the inclusion of particle reactions is an interesting direction for future work. Also, the incorporation of time-variant flows to model, e.g., the pumping of blood, and the analysis of more complex RX and TX models are of interest.  




%



%
%

\ifCLASSOPTIONcaptionsoff
  \newpage
\fi



\bibliographystyle{IEEEtran}
\bibliography{IEEEabrv,%
	./tmbmc.bib}
\end{document}